\documentclass[twocolumn]{aastex631}

\usepackage{graphicx}
\usepackage{amsmath}	
\usepackage{amssymb}	
\usepackage{float}
\usepackage{subfigure}
\usepackage{multirow}
\usepackage{enumerate}
\usepackage{color}
\usepackage{amsmath}
\usepackage{tabularx}
\usepackage{booktabs}
\usepackage{threeparttable}
\usepackage[utf8]{inputenc}
\usepackage{relsize}
\usepackage{pgffor}
\usepackage{filecontents}
\usepackage{subfigure}
\usepackage[mathlines]{lineno}
\usepackage{booktabs}
\usepackage{ragged2e}

\begin{document}


\title{Revisiting open clusters within 200\,pc in the solar neighbourhood with Gaia DR3}

\correspondingauthor{Min Fang}
\email{mfang@pmo.ac.cn}

\author{Penghui Liu}
\affiliation{Purple Mountain Observatory, Chinese Academy of Sciences, 10 Yuanhua Road, Nanjing 210023, People's Republic of China
}
\affiliation{
    School of Astronomy and Space Science, University of Science and Technology of China, Hefei 230026, China;
}

\author{Min Fang}
\affiliation{Purple Mountain Observatory, Chinese Academy of Sciences, 10 Yuanhua Road, Nanjing 210023, People's Republic of China
}
\affiliation{
    School of Astronomy and Space Science, University of Science and Technology of China, Hefei 230026, China;
}

\author{Yue-Lin Sming Tsai}
\affiliation{Purple Mountain Observatory, Chinese Academy of Sciences, 10 Yuanhua Road, Nanjing 210023, People's Republic of China
}
\affiliation{
    School of Astronomy and Space Science, University of Science and Technology of China, Hefei 230026, China;
}

\author{Xiaoying Pang}
\affiliation{Department of Physics, Xi'an Jiaotong-Liverpool University, 111 Ren'ai Road, Dushu Lake Science and Education Innovation District, Suzhou 215123, Jiangsu Province, People's Republic of China
}
\affiliation{
Shanghai Key Laboratory for Astrophysics, Shanghai Normal University, 100 Guilin Road, Shanghai 200234, People's Republic of China
}

\author{Fan Wang}
\affiliation{Purple Mountain Observatory, Chinese Academy of Sciences, 10 Yuanhua Road, Nanjing 210023, People's Republic of China
}
\affiliation{
    School of Astronomy and Space Science, University of Science and Technology of China, Hefei 230026, China;
}

\author{Xiaoting Fu}
\affiliation{Purple Mountain Observatory, Chinese Academy of Sciences, 10 Yuanhua Road, Nanjing 210023, People's Republic of China
}
\affiliation{
    School of Astronomy and Space Science, University of Science and Technology of China, Hefei 230026, China;
}

\begin{abstract}

In this study, we develop a membership identification method and apply it for 30 open clusters (OCs) within 200~pc of the Sun using astrometric data of Gaia DR3. 
By accounting for projection effects that distort apparent stellar motions, our approach converts astrometric data into accurate five-dimensional positions and velocities. 
This approach enables better identification of members in nearby open clusters. We then compare our refined membership lists with previous catalogs, revealing more members in most open clusters, but also the identification of elongated structures in Melotte~25 (Hyades), NGC~2632 (Praesepe), Melotte~111 (Coma Berenices), Platais~3, Melotte~22 (Pleiades), NGC~2451A,Platais~9, IC~2391, Platais~8, UPK~640, HSC~2986, which we studied in detail. An analysis of the ages of their members reveals the members within and outside of the tidal radius are distinctly coeval,  further validating our methodology. This study suggests that for OCs in the solar neighborhood, correcting for the projection effect is very important for identification of OC members.

\end{abstract}


\section{Introduction}\label{sec:in}
Open clusters (OCs) are an ideal place to study the Galactic structure and theories of stellar evolution. The members of the OCs were born almost simultaneously in the same molecular cloud~\citep{lada2003embedded}, they are spatially close and have similar velocities, which gives us clues to find them. The latest Gaia data release 3~\citep{vallenari2023gaia} provide unprecedented accuracy in astrometric measurements of sources, including radial velocity measurements for more than 33 million sources, helping to discover OCs and identify their potential members. 
Since the release of data from the Gaia satellite~\citep{prusti2016gaia, gaiadr12016, gaiadr22018, gaiaedr32021}, 
machine-learning based methods have been widely adopted for identifying OCs and their members. 
Utilizing machine learning methods to search for OCs within the expanding dataset provided by Gaia yields reliable results and is instrumental in identifying OC members~\citep[see e.g.][]{castro2018new}. A common way to search for clusters is to use five-dimensional astrometric data in combination with clustering algorithms such as FoF \citep{davis1985evolution}, DBSCAN \citep{1996A}, or HDBSCAN \citep{2017hdbscan} to search for OCs by dividing the data into different regions~\citep[see e.g.][]{liu2019catalog, castro2020hunting, castro2022hunting, hunt2021improving, he2022unveiling}. 

Being closer to the Sun enables more accurate distance measurements of OCs and helps us better understand their properties. Studying the morphology and structure of nearby OCs provides valuable clues about their formation and dynamics, including three-dimensional shapes and tidal tails~\citep{meingast2021extended, tarricq_structural_2022}. 
In general, nearby OCs have small distance modulus and low extinction, making it possible to  study their population till the very low mass part to advance our understanding of star formation and evolution~\citep{paradis2012dark, capitanio2017three, babusiaux2018gaia}. 
On the other hand, due to internal dynamics in OCs and external tidal disruption of the Milky Way, a small fraction of nearby OCs have pronounced tidal tails \citep{roser2019hyades,ye2021extended,kroupa2022asymmetrical}. Studying the tidal tails not only help to constrain the evolution and disruption of these clusters, but also is crucial for our understanding of the Milky Way cluster formation history~\citep{lamers2005analytical,yeh2019ruprecht,tarricq_structural_2022}. Improved data quality and algorithmic advances have enabled the discovery of many new stellar groups in space and velocity space, even within the well-studied solar neighborhood~\citep{moranta2022new,zucker2023solar,cantat2024gaia}. Besides discovering nearby stellar groups, it is important to comprehensively characterize their members, such as tidal tails and kinematic properties~\citep{roser2019hyades, tang2019discovery, meingast2019extended, meingast2021extended, 2021ApJ...912..162P}. Particularly in the Gaia era, which has witnessed a surge in the identification of stellar groups, there remains a need for more detailed analyses of their constituent members~\citep{perren2023unified}.

The projection effect is prominent in nearby OCs, as stars moving relative to the observer appear to have different tangential velocities depending on their angular separation from the cluster center \citep{van2009parallaxes}. Even if the members of a nearby OC have exactly the same velocity, their proper motions will vary due to the projection effect. Therefore, it is necessary to consider the projection effect and make corrections when determining the members of nearby OCs.

In this study, we present a membership determination method specifically designed for nearby open clusters from \cite{hunt2023improving} (hereafter Hunt~23) within 200~pc. Our approach utilizes Gaia DR3 data and corrects for projection effect on stellar motions by transforming the astrometric parameters into three-dimensional positions and velocities relative to the bulk motion of each cluster. We apply this technique to a sample of 30 nearby open clusters and compare our results with previous membership catalogs. Our main goals are to expand membership lists, identify extended structures and tidal tails that reflect cluster dissolution, and gain insights into their formation and evolution.

The structure of this paper is as follows. In Sec.~\ref{sec:sad}, we present the selected OCs, the selection and screening criteria of Gaia DR3 sources around the OCs. In Sec.~\ref{sec:mm}, we introduce our membership determination method for open clusters, focusing on the application of the HDBSCAN algorithm and our approach to correcting projection effect. In Sec.~\ref{subsec:rad}, we present the results of our membership analysis, comparing our findings with previous catalogs and highlighting the discovery of new members and the properties of these OCs based on their members. We summarized this work in Sec.~\ref{sec:con}.

\section{Sample and Data}\label{sec:sad}
In this section, we briefly describe the selection criteria for the open cluster sample and Gaia DR3 sources and how we preprocess these data.

\subsection{OCs sample}\label{subsec:os}

We identify cluster member stars based on the OC catalog from Hunt~23 and limit our cluster selection within 200~pc of the Sun. Hunt~23 performed a blind all-sky search for open clusters based on the latest Gaia DR3 data~\citep{vallenari2023gaia}, using an improved open cluster census approach. For our chosen range of distances, OCs are strongly affected by the projection effect (see Sec.~\ref{subsubsec:pe}), and we expect to find more members using our method, making the cluster membership list more complete, and giving a better understanding of their member composition. We select only those labeled "\textit{o}" (the open clusters) but remove those labeled "\textit{m}" (the moving groups) in the catalog. 
We remove the OCs Chamaleon~I and HSC~2919 because they are relatively young and located in star forming regions, resulting in poor performance of the reddening correction method described in Sec~\ref{subsec:esoo}. Note that we selected only those OCs with a number of members greater than 100, as such clusters are statistically significant. In total we have 30 OCs in the sample. Among them, 6 OCs have been also studied in \cite{tarricq_structural_2022} (hereafter Tar~22), and 13 OCs investigated in  \cite{cantat2020painting}(hereafter CG~20). The parameters of our studied OCs are presented in Table.~\ref{tab1}.

\begin{table*}[htbp]
\centering
\caption{The selected OCs in this Work.}
\label{tab1}
{\fontsize{30}{35}\selectfont
\resizebox{1.\textwidth}{!}{
{
\begin{tabular}{l c c c c c c c | c c c c}
    \hline
    \textbf{Name} & \textbf{Num} & \textbf{Ra} & \textbf{Dec} & \textbf{PMRA} & \textbf{PMDEC} & \textbf{Distance} & \textbf{Age} & \textbf{Num} & \textbf{Elongated} & \textbf{Age} & \textbf{References} \\ 
     ~ & (Hunt~23) & (deg) & (deg) & (mas $yr^{-1}$) & (mas $yr^{-1}$) & (pc) & (log(yr)) & (this study) & ~ & (this study) & ~  \\ 
    \hline
Alessi~13   & 167 & 51.99  & -35.77 & 35.50  & -4.18  & 104.25 & $7.39^{+0.19}_{-0.34}$ & 144 & No & $7.63^{+0.01}_{-0.02}$ & 1,3 \\
Alessi~84   & 128 & 110.51 & 55.38  & -3.12  & -31.45 & 198.30 & $8.07^{+0.25}_{-0.25}$ & 1069 & No & $8.57^{+0.08}_{-0.02}$ & 1 \\
HSC~396     & 134 & 317.19 & -3.67  & 21.80  & -8.72  & 99.14  & $8.13^{+0.28}_{-0.37}$ & 110 & No & $8.65^{+0.06}_{-0.07}$ & 1 \\
HSC~749     & 146 & 358.47 & 3.46   & 18.06  & -13.84 & 121.15 & $8.35^{+0.30}_{-0.41}$ & 240 & No & $8.19^{+0.02}_{-0.04}$ & 1 \\
HSC~759     & 372 & 225.15 & 59.39  & -16.19 & -3.64  & 95.96  & $8.17^{+0.27}_{-0.29}$ & 283 & No & $8.75^{+0.01}_{-0.04}$ & 1 \\
HSC~976     & 235 & 10.00  & 79.32  & 17.16  & 5.32   & 163.08 & $7.13^{+0.26}_{-0.21}$ & 311 & No & $7.42^{+0.01}_{-0.01}$ & 1 \\
HSC~1340    & 194 & 61.30  & 21.99  & 0.72   & -14.55 & 120.09 & $7.42^{+0.25}_{-0.26}$ & 180 & No & $7.56^{+0.01}_{-0.06}$ & 1 \\
HSC~1438    & 110 & 60.27  & 8.34   & 24.00  & -23.57 & 147.62 & $7.75^{+0.25}_{-0.23}$ & 94 & No & $7.81^{+0.02}_{-0.03}$ & 1 \\
HSC~1542    & 175 & 93.72  & 15.55  & 0.84   & -10.59 & 152.98 & $8.25^{+0.28}_{-0.26}$ & 87 & No & $8.14^{+0.19}_{-0.17}$ & 1 \\
HSC~2303    & 153 & 169.14 & -35.60 & -42.07 & -2.20  & 154.64 & $7.90^{+0.24}_{-0.29}$ & 169 & No & $8.39^{+0.04}_{-0.03}$ & 1 \\
HSC~2468    & 210 & 182.23 & -51.54 & -35.27 & -11.96 & 108.98 & $6.92^{+0.17}_{-0.28}$ & 222 & No & $7.32^{+0.02}_{-0.02}$ & 1 \\
HSC~2505    & 119 & 185.02 & -64.10 & -37.79 & -10.81 & 105.85 & $6.84^{+0.17}_{-0.21}$ & 123 & No & $7.19^{+0.02}_{-0.02}$ & 1 \\
HSC~2636    & 284 & 204.89 & -44.48 & -25.97 & -19.41 & 133.88 & $6.99^{+0.23}_{-0.25}$ & 321 & No & $7.36^{+0.01}_{-0.02}$ & 1 \\
HSC~2733    & 165 & 228.06 & -44.53 & -20.67 & -21.89 & 144.17 & $6.98^{+0.20}_{-0.25}$ & 141 & No & $7.43^{+0.03}_{-0.01}$ & 1 \\
HSC~2907    & 349 & 244.77 & -24.45 & -10.68 & -21.70 & 154.58 & $6.99^{+0.19}_{-0.27}$ & 284 & No & $7.18^{+0.0}_{-0.03}$ & 1 \\
HSC~2986    & 222 & 285.32 & -36.88 & 2.21   & -27.56 & 148.05 & $6.85^{+0.19}_{-0.21}$ & 475 & Yes & $7.13^{+0.05}_{-0.01}$ & 1 \\
IC~2391     & 376 & 130.26 & -53.04 & -24.80 & 23.27  & 150.19 & $7.44^{+0.24}_{-0.28}$ & 607 & Yes & $7.74^{+0.01}_{-0.01}$ & 1,3 \\
IC~2602     & 638 & 160.97 & -64.39 & -17.78 & 10.67  & 150.55 & $7.41^{+0.26}_{-0.23}$ & 583 & No & $7.70^{+0.01}_{-0.03}$ & 1,3 \\
Melotte~20  & 938 & 51.33  & 49.12  & 22.92  & -25.46 & 173.58 & $7.75^{+0.23}_{-0.27}$ & 2544 & No & $7.86^{+0.0}_{-0.01}$ & 1,2,3 \\
Melotte~22 (Pleiades)  & 1721 & 56.68  & 24.11  & 19.96  & -45.46 & 134.84 & $8.08^{+0.21}_{-0.34}$ & 1763 & Yes & $8.09^{+0.01}_{-0.01}$ & 1,2,3 \\
Melotte~25 (Hyades)  & 927 & 66.71  & 16.08  & 104.14 & -28.73 & 47.19  & $8.76^{+0.23}_{-0.21}$ & 976 & Yes & $8.89^{+0.01}_{-0.07}$ & 1,3 \\
Melotte~111 (Coma Berenices) & 271 & 186.02 & 26.42  & -12.15 & -8.86  & 85.25  & $8.82^{+0.22}_{-0.19}$ & 360 & Yes & $8.81^{+0.01}_{-0.01}$ & 1,2,3 \\
NGC~2632 (Praesepe)    & 1314 & 130.09 & 19.67  & -35.94 & -12.90 & 183.49 & $8.54^{+0.25}_{-0.23}$ & 1629 & Yes & $8.86^{+0.01}_{-0.01}$ & 1,2,3 \\
NGC~2451A   & 407 & 115.97 & -38.24 & -20.99 & 15.33  & 190.08 & $7.42^{+0.27}_{-0.30}$ & 1239 & Yes & $7.83^{+0.01}_{-0.01}$ & 1,3 \\
OCSN~92     & 103 & 231.56 & -36.06 & -20.50 & -24.58 & 138.17 & $6.97^{+0.25}_{-0.21}$ & 129 & No & $7.40^{+0.02}_{-0.03}$ & 1 \\
OCSN~96     & 223 & 240.43 & -22.71 & -11.54 & -23.89 & 140.61 & $6.67^{+0.12}_{-0.16}$ & 213 & No & $6.99^{+0.01}_{-0.03}$ & 1 \\
Platais~3   & 204 & 68.41  & 71.49  & 3.77   & -20.68 & 177.21 & $8.24^{+0.24}_{-0.30}$ & 450 & Yes & $8.71^{+0.01}_{-0.07}$ & 1,2,3 \\
Platais~8   & 225 & 136.90 & -59.31 & -16.24 & 13.74  & 133.59 & $7.49^{+0.33}_{-0.28}$ & 673 & Yes & $7.63^{+0.01}_{-0.01}$ & 1,3 \\
Platais~9   & 251 & 138.32 & -43.56 & -24.56 & 13.30  & 185.26 & $7.57^{+0.25}_{-0.26}$ & 364 & Yes & $7.83^{+0.02}_{-0.01}$ & 1,2,3 \\
UPK~640     & 764 & 249.69 & -39.51 & -11.88 & -21.39 & 173.86 & $6.98^{+0.18}_{-0.26}$ & 779 & Yes & $7.32^{+0.01}_{-0.01}$ & 1,3 \\
\bottomrule

\end{tabular}
}}
\\
\begin{tablenotes}
\footnotesize
\item{\textbf{Note:}  Basic parameters of the selected OCs. The left side of the split line shows the basic parameters of the 30 OCs we selected in Hunt~23. The right side of the split line shows the properties of these OCs obtained from this study, where elongated indicates whether the elongated morphology of the clusters can be seen based on our membership, and references indicates the work to which our membership has been compared. On both sides, the 16th and 84th percentiles of the logarithm of cluster age are indicated.}
\end{tablenotes}

\begin{tablenotes}
\footnotesize
\item{\textbf{Reference:} (1) Hunt~23; (2) Tar~22; (3) CG~20.}
\end{tablenotes}}
\end{table*}

\subsection{Data from Gaia DR3}\label{subsec:dfgd}
We use the latest release of Gaia DR3 and select sources with five-dimensional astrometric $(\alpha, \delta, \varpi, \mu_ {\alpha^{\ast}}, \mu_ {\delta})$ and three-dimensional photometric $(G, G_{BP}, G_{RP})$. For nearby regions of the sky, the members of an OC will typically cover a large area of the sky region, ranging from a few degrees to several tens of degrees, the same situation for \texttt{proper motions}. We refer to the method in Tar~22 and require a cone of $70$~pc radius around the center of the cluster based on the mean \texttt{sky position}, \texttt{proper motions}, and \texttt{parallax} of the cluster as provided by Hunt~23. We note that for clusters with radii larger than $70$~pc in Hunt~23, HSC~749, HSC~759, and HSC~2303, we extended the cone radius to $100$~pc to ensure that the membership analysis fully covers their regions. Furthermore, we only take those stars within $15$ standard deviation of the cluster mean \texttt{proper motions} and $1000/{\varpi}<1000/{\varpi_{c}} + 200 \ pc$, where $\varpi_{c}$ is the mean \texttt{parallax} of that OC. This aims to broaden the coverage of candidate stars in order to identify potential OC members that might have been overlooked due to the projection effects discussed in Sec.~\ref{subsubsec:pe}.

\subsection{Unreliable source  removal}\label{subsec:rus}

Gaia DR3 offers high-quality measurements, but some of them do not reflect their true values, particularly for sources near the Sun. 
For instance, approximately half of the sources located within 100 pc from the Sun are spurious~\citep[hereafter Ryb~22]{rybizki2022classifier}. 
Thus, when using Gaia DR3 data in the solar neighborhood, we need to filter out these spurious data.

To filter out the spurious sources, we use the astrometric fidelity condition ($\texttt{v2}\ge 0.5$) as given in Ryb~22, instead of a simple Renormalised Unit Weight Error (\texttt{RUWE}) or magnitude cuts. According to \cite{hunt2023improving}, utilizing the astrometric spurious quality flag alone for cluster reconstruction can be better than using either \texttt{RUWE} or magnitude cuts. This methodology enhances cluster member identification by increasing confirmed cluster members, reducing spurious detections, and improving discrimination between cluster and field populations.

We also test data filtering criteria: (i) $\texttt{parallax\_over\_error} > 10$ OR $\texttt{parallax\_error} < 0.1$, and (ii) $\texttt{astrometric\_excess\_noise} < 1$ OR $2> \texttt{astrometric\_excess\_noise} > 1$, as proposed by \cite{kounkel2019untangling}. We find that these criteria have filtered out a large number of sources that are reliable according to the criteria of Ryb~22. Taking IC~2602, an OC at around 150~pc far from the Sun, as an example, filtering using the fidelity criterion yields 13\% more candidates data than that obtained by the above-mentioned criteria. In addition, clustering on the data selected by the fidelity criterion, when using HDBSCAN, yields 3\% more corresponding cluster members than the above-mentioned filtering criteria.

\section{Membership methodology}\label{sec:mm}

\subsection{HDBSCAN}\label{subsec:hd}

In this work we employ the HDBSCAN \citep{2017hdbscan} algorithm to identify members of the selected OCs.

HDBSCAN is a density-based clustering algorithm that constructs a hierarchy-based tree to cluster data points. It improves and extends the DBSCAN \citep{1996A} algorithm. 
The algorithm has the following features:

\begin{itemize}
    \item It is adapted to the optimal clustering result based on the distribution of data.
    \item Unlike DBSCAN, which is applicable to the same density cases, HDBSCAN can also find clusters of different densities. 
    \item It can use hierarchical clustering to extract clusters.
\end{itemize}

HDBSCAN has two parameters, $m_{\rm clSIZE}$ and $m_{\rm Pts}$, which are typically set to the same value. The parameter $m_{\rm clSIZE}$ represents the minimum number of cluster members. If the number of cluster members is less than $m_{\rm clSIZE}$, it is considered noisy. The parameter $m_{\rm Pts}$ represents the minimum number of points around a core point, used for the confirmation of cluster boundaries in the clustering process. For the cluster selection method, we follow~\cite{hunt2021improving} and choose the leaf method. The excess of mass method, in contrast, often groups the sources of the whole region into one or two large groups, thus it is not suitable to identify the members of OC \citep{kounkel2019untangling}.

\subsection{Correction}\label{subsec:co}

This section introduces the concept of the projection effect and its application in data processing for determining OC membership.

\subsubsection{Projection effect}\label{subsubsec:pe}

\begin{figure*}
    \centering
    \includegraphics[width=0.7\linewidth]{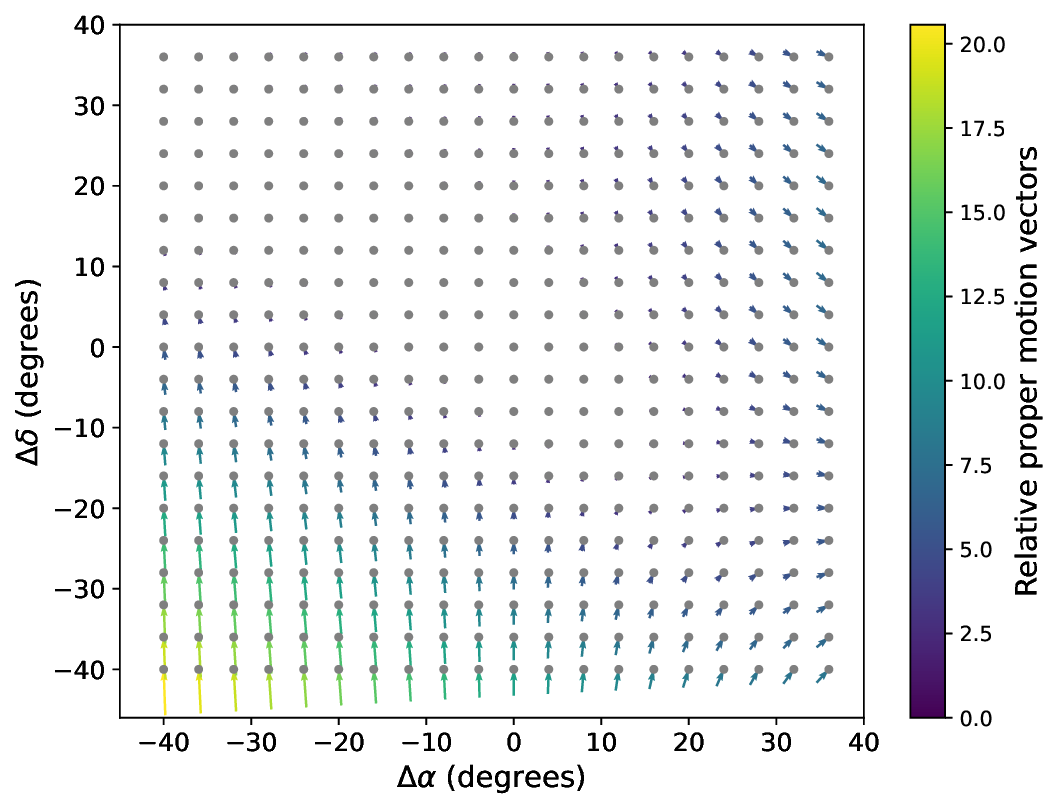}
    \caption{UPK~640 as a demonstrative case to highlight the variations in \texttt{proper motion} due to projection effect, assuming that all of its members have a common three-dimensional velocity (based on the mean value of the Hunt~23 members). 
    By utilizing the cluster aggregate motion, the relative \texttt{proper motion} vectors at various positions in the sky relative to the cluster center is presented with equal distances from the Sun. 
    The $\Delta\alpha$ and $\Delta\delta$ represent the angle of deviation from the cluster center in the directions of \texttt{ra} and \texttt{dec}, respectively. 
    The length and direction of the color arrows show the relative vectors with respect to the gray dots (the selected sky coordinates).
    }
    \label{fig:projection}
\end{figure*}

The projection effect arises because apparent stellar motions in nearby clusters, as seen from Earth, are distorted by perspective. Stars moving radially outward or inward relative to the observer appear to have different tangential velocities depending on their angular separation from the cluster center. Our correction method transforms observed proper motions into a five-dimensional framework that decouples intrinsic stellar motions from projection-induced distortions.

In an area closed to the Sun, the members of the OC can be scattered widely in the sky. Even if the members of an OC have exactly the same three-dimensional velocity, their \texttt{proper motions} will be different due to projection effect. In Fig.~\ref{fig:projection}, we demonstrate the relative values of proper motion (colored vectors) projected on the $(\Delta \alpha, \Delta \delta)$ plane with cluster UPK~640 as an example. Parameters of this cluster are listed in the last row of Table.~\ref{tab1}. The values of $\Delta \alpha$ and $\Delta \delta$ represent the angle difference from the cluster center in the directions of \texttt{ra} and \texttt{dec}, respectively. The gray dots presents the selected sky coordinates, 
while the arrows are these relative vectors with illustrated strengths (lengths and colors) and directions. We can clearly see that the \texttt{proper motion} corresponding to the positions far from the center can significantly differ from that at the center of the cluster. Compared to clusters of the same size farther away, due to their large distribution across the sky, clusters within 200~pc of the Sun have a much more pronounced projection effect.

To mitigate the projection effect, it is necessary to consider the total motion of the OC. Our correction strategy is to calculate the mean values of six parameters for an OC: $(\alpha, \delta, \varpi, \mu_{\alpha^{\ast}}, \mu_{\delta})$, and $v_{\rm rad}$. 
These values are used to represent the overall spatial distribution and motion of the cluster. 
We refer to~\cite{van2009parallaxes} for more details. The correction matrix is 

\begin{equation}
\left[\begin{array}{l}V_{\mathrm{rad},i}\\
\kappa\,\frac{\mu_{\alpha*,i}}{\varpi_i} \\ 
\kappa\,\frac{\mu_{\delta,i}}{\varpi_i} 
\\\end{array}\right] = 
\left[\begin{array}{rrr} 
\cos\alpha_i\cos\delta_i & \sin\alpha_i\cos\delta_i & \sin\delta_i \\
-\sin\alpha_i & \cos\alpha_i & 0 \\ 
-\cos\alpha_i\sin\delta_i & -\sin\alpha_i\sin\delta_i & \cos\delta_i \\
\end{array}\right]
\,\cdot\,\dot{\boldsymbol{R}},
\label{equ:1}
\end{equation}

where we define $\boldsymbol{v}_{i} \equiv (V_\mathrm{rad,i},\kappa\mu_{\alpha*,i}/\varpi_{i},\kappa\mu_{\delta,i}/\varpi_{i})$, and the rotation matrix on the right hand side is $\boldsymbol{A_i}$. 
The velocity vector of the cluster with respect to the Sun is defined as $\dot{\boldsymbol{R}} = {\boldsymbol{A}}^{-1}_0\,\boldsymbol{v}_0$, thus we have

\begin{equation}
\boldsymbol{v}_i = \boldsymbol{A}_i \dot{\boldsymbol{R}} = \boldsymbol{A}_{i} \boldsymbol{A}^{-1}_0\,\boldsymbol{v}_0.
\label{equ:2}
\end{equation} 

The subscript $i$ refers to different positions of each individual cluster member star, while the subscript $0$ refers to the cluster itself. Following \cite{van2009parallaxes}, we have

\begin{eqnarray}
\Delta\mu_{\alpha *,i} = \mu_{\alpha*,i}\,\frac{\varpi_0}{\varpi_i} - \mu_{\alpha,0}, \nonumber\\  
\Delta\mu_{\delta,i} = \mu_{\delta,i}\,\frac{\varpi_0}{\varpi_i} - \mu_{\delta,0}.
\label{equ:3}
\end{eqnarray}

At different \texttt{sky positions}, even if all the member stars of an OC have the same three-dimensional velocity, their apparent proper motions will differ from each other due to the projection effect.

Taking into account the distances of various candidate stars, we can derive the residual proper motion of a candidate star from the cluster, 

\begin{eqnarray}
R_{ \alpha * , i } = \texttt{PMRA}_{i} -\left(\Delta \mu _ { \alpha * , i } + \mu _ { \alpha , 0 }\right )\times  \frac{d_0}{d_i}, \\
R_{ \delta, i } = \texttt{PMDEC}_{i} -\left(\Delta \mu _ { \delta , i } + \mu _ { \delta, 0 }\right) \times \frac{d_0}{d_i}, 
\label{equ:4}
\end{eqnarray}

where the proper motions in the corresponding directions of the candidate star denote as $\texttt{PMRA}_i$ and $\texttt{PMDEC}_i$. The distance from the cluster and candidate star to the Sun is denoted as $d_0$ and $d_i$.

\subsubsection{Radial velocities}\label{subsubsec:rv}

To obtain the residual proper motion of a candidate star as mentioned in Sec.~\ref{subsubsec:pe}, we need an estimate of the overall motion of the cluster (as described in Eq.~\ref{equ:1} and Eq.~\ref{equ:2}), which requires estimating the radial velocities of the cluster. Gaia DR3 provides high-precision \texttt{radial velocity} (\texttt{RV}) measurements for over 33 million stars, approximately four times more than Gaia DR2. For each selected cluster, we use the data, \texttt{RV} and \texttt{RV error}, of its members to estimate the \texttt{RV} of the whole OC system.~\footnote{Around one-third of the OC members are with \texttt{RV} data available.}
To address this, we propose a method as following. 
We initially select data by applying the criterion $\texttt{RV~error}<3~\textnormal{km s}^{-1}$ to eliminate sources with large errors, thus ensuring accurate \texttt{RV} measurements for better cluster \texttt{RV} estimation. However, due to projection effects, even perfectly accurate \texttt{RV} measurements of cluster members can show large differences. Subsequently, referring to \cite{bhattacharya2022gaia}, we calculate the median and standard deviation of the \texttt{RV} estimates for the remaining sources. We then remove those that fall outside the $2\sigma$ region of the median. The \texttt{RV} distributions of the remaining members are typically much closer. Finally, we compute the mean \texttt{RV} and standard deviation of these remaining members, which we take as representing the overall \texttt{RV} distribution of the cluster system.

\subsubsection{Clustering}\label{subsubsec:cl}
 
For each OC, we estimate the \texttt{RV} and its standard deviation. We then utilize the mean \texttt{sky position} and \texttt{proper motion} data of the OC from Hunt~23. For each cluster, we sample the \texttt{RV} based on the estimated value and standard deviation, subsequently correcting the queried data using the method outlined in Sec.~\ref{subsubsec:pe}. Additionally, high-latitude spherical distortions present a challenge, which is addressed by combining the source sky position with its \texttt{photogeometric distances} from \cite{bailer2021estimating} and converting the data to \texttt{heliocentric Cartesian coordinates}.

After combining three-dimensional \texttt{heliocentric Cartesian coordinates} and \texttt{proper motion}, we obtain a new five-dimensional coordinate, 

\begin{equation*}
    \left(x, y, z, R_{ \alpha * , i }, R_{ \delta, i }\right),
\end{equation*}

where $x$, $y$, and $z$ are the corresponding \texttt{heliocentric Cartesian coordinates}. Before the clustering process, $R_{ \alpha * , i }$ and $ R_{ \delta, i }$ are converted to the units of km~s$^{-1}$ using the factor $\kappa = 4.74047$, which converts proper motion in mas $yr^{-1}$ to velocity in km~s$^{-1}$ at a distance of 1 kpc. The method \textbf{RobustScaler}~\citep{pedregosa2011scikit} is used to rescale the five-dimensional data. It reduces the effect of outliers compared to standard normalization because it uses medians and quartiles instead of means and variances for the normalization process \citep{hunt2023improving}.

In HDBSCAN setup, we take the minimum cluster size $m_{\rm clSIZE}=80$ and the minimum number of points around a core point $m_{\rm Pts}=10$. 
For each cluster, we generate 100 samples of the \texttt{RV} based on its estimated mean and standard deviation. Using each sampled \texttt{RV}, we apply the correction method mentioned in Sec.~\ref{subsubsec:pe} to the queried data, producing a set of parameters $(x, y, z, R_{ \alpha * , i }, R_{ \delta, i })$. HDBSCAN then performs clustering on the corrected data in this five-dimensional space. Finally, the membership probability of a star belonging to the cluster is the ratio of times identified as a cluster member to the total of 100 samples.

\section{Result and discussion}\label{subsec:rad}

\subsection{Comparison with previous studies}\label{subsec:moe}

\begin{figure*}
    \centering
    \includegraphics[width=1.\linewidth]{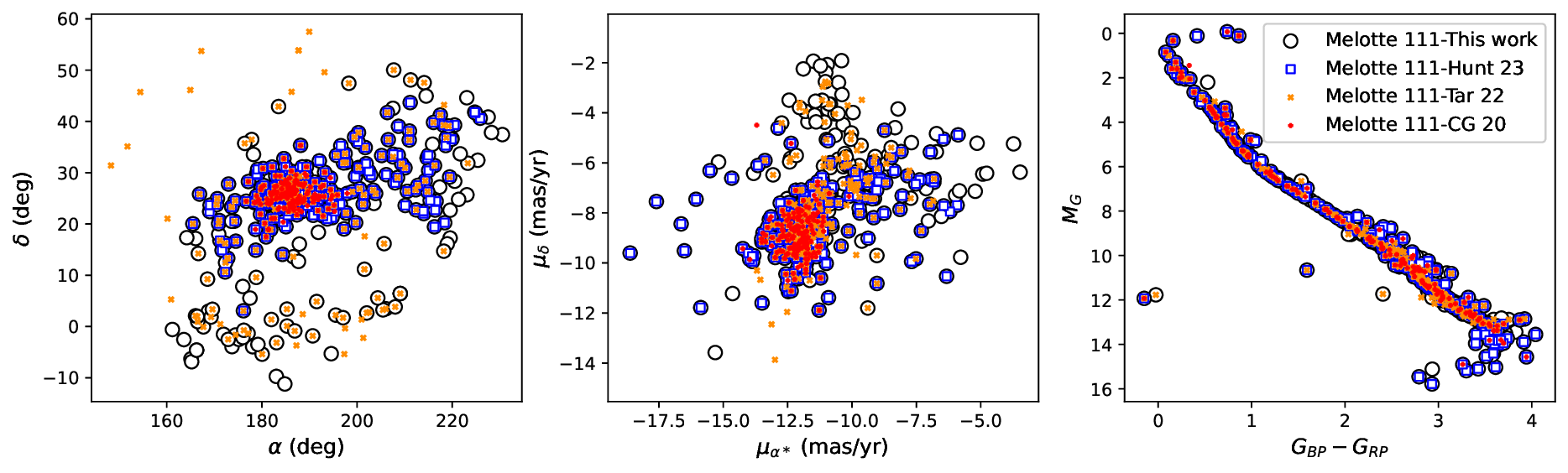}
    \caption{Comparison between our members and the members of Hunt~23, Tar~22 and CG~20 in Melotte~111.}
    \label{fig:compare dif member}
\end{figure*}

In this subsection, we compare the OC members identified by our method with those from Hunt~23, Tar~22, and CG~20. As an example, we present the distribution of stars in Melotte~111 projected onto three different planes in Fig.~\ref{fig:compare dif member}. 
Our results are marked with black circles, while those from Hunt~23, Tar~22, and CG~20 are shown as blue squares, orange crosses, and red solid dots, respectively, for comparison. We discuss their differences as follows.

\textbf{\underline{Hunt~23}:}

We compare the members in this study with those identified by Hunt~23, which uses the HDBSCAN method. They optimize member determination for the nearby sky region within 250~pc by using three-dimensional spatial coordinates instead of the sky coordinates and distances. We retain only our members with a membership probability greater than 0.5. 
This selection criterion is a consequence of our approach to sampling the \texttt{RVs} of OCs and applying corrections to the queried data as described in Sec.~\ref{subsubsec:cl}. By generating 100 RV samples for each cluster and performing clustering in the corrected parameter space, we ensure that only members with high membership probabilities (greater than 0.5) are retained to minimize contamination. 
Members with low probabilities are susceptible to contamination and are therefore excluded from the membership list. The distributions of members in terms of sky positions, proper motions, and color-magnitude diagrams (CMD) for all OCs\footnote{All members identified in this study are available in their entirety in a machine-readable form.} are presented in Appendix \ref{app:doo}.

For most of the OCs, we have more members than Hunt~23 provides, some of which are only 10\% more than Hunt~23 (e.g., OCSN~92), while some of them reach multiples of Hunt~23 (e.g., Platais~8). For some OCs, we have almost the same number of members as Hunt~23 (e.g., UPK~640). 
Our method effectively detects the elongated morphology of nearby open clusters, while typically identifying more cluster members than Hunt~23. 
For example, as we can see from Fig.~\ref{fig:compare dif member} and Table.~\ref{tab1} that Hunt~23 detects 271 members for Melotte~111, while our method can detects 360 members. Both methods are with significant tidal tails. Our relatively higher numbers of members in our studied OCs are likely due to the correction for the projection effect of \texttt{proper motions} in our method. This allows us to obtain more members outside the OC core that are influenced by the projection effect.

\textbf{\underline{Tar~22}:}

We evaluated our approach against the OCs of the membership list provided by Tar~22. 
Tar~22 has utilized the HDBSCAN algorithm to construct their membership lists. They apply HDBSCAN to the Gaia EDR3 \texttt{parallax} and \texttt{proper motion} dimensions $(\varpi, \mu_{\alpha^{\ast}}, \mu_{\delta})$ without additional selection criteria in the \texttt{sky position} dimensions. Tar~22 focused on the morphological properties and mass separation levels of clusters, thus performing HDBSCAN only in parallax-proper motion space to avoid affecting stars in the outskirts of the clusters. However, their approach generates successful membership lists for only 389 out of 467 OCs, as clusters with overlapping neighbors in the same field and close parallax-proper motion space were identified, leading to unsuccessful memberships. 
By converting mas~yr$^{-1}$ to km~s$^{-1}$ to avoid variations due to the distance of neighboring celestial regions, our membership results are more reasonable and refined.

For the OCs chosen by our approach and Tar~22, we always find more members than Tar~22. For OCs in close proximity, members exhibit a wide range of \texttt{proper motions}. 
Compared to our approach, the method used by Tar~22 identifies members with a narrower range of \texttt{proper motions}. 
For instance, in the Platais~3 cluster, approximately 170~pc from the Sun, we identified members with ranges, $-11\le \mu_{\alpha^{\ast}}/({\rm mas~yr^{-1}})\le 12$ and $-27\le\mu_{\delta}/({\rm mas~yr^{-1}})\le -15$. 
In contrast, Tar~22 method finds members with ranges, $0\le\mu_{\alpha^{\ast}}/({\rm mas~yr^{-1}})\le 7$ and $-23\le \mu_{\delta}/({\rm mas~yr^{-1}})\le -18$. 
Our approach, which uses 3D spatial coordinates and corrected projected velocities, results in a membership list with a broader range of \texttt{proper motions} compared to that of Tar~22.

\textbf{\underline{CG~20}:}

\begin{figure*}
    \centering
    \includegraphics[width=1.\linewidth]{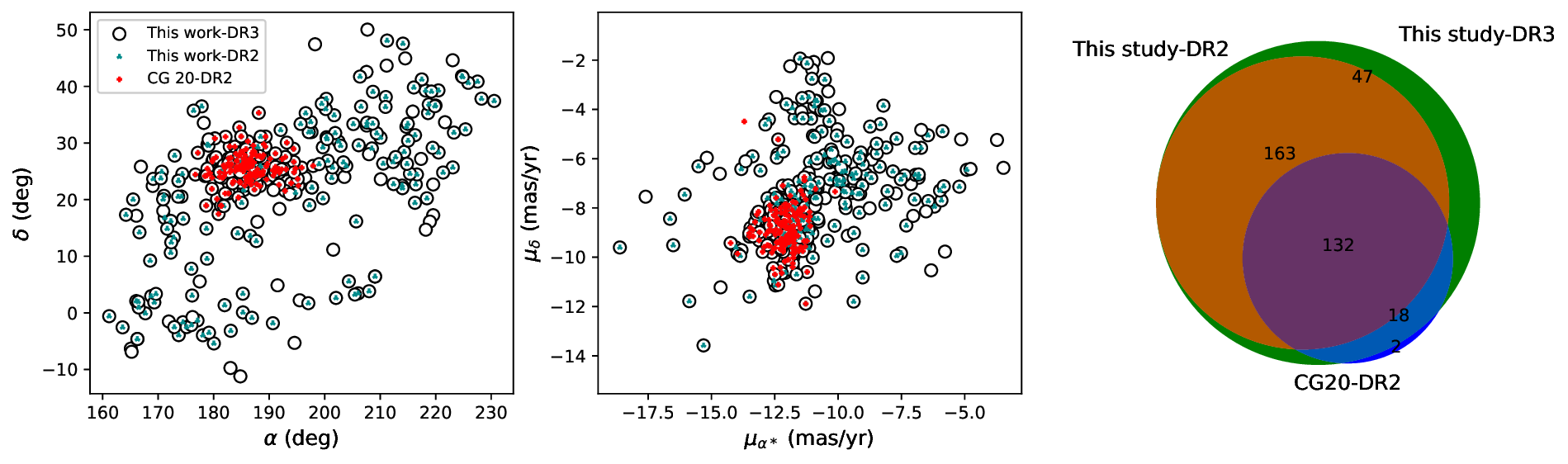}
    \caption{Comparison between our members (Gaia DR2 and DR3) and the members of CG~20 in Melotte 111. The left and middle panels display the distribution of members in sky position and proper motion respectively. The right panel presents a Venn diagram of the membership overlap. The intersection of members from this study (DR2) and CG~20 (DR2) is shown in purple, while members unique to this study (DR3) are highlighted in green. The remaining regions follow similarly.}
    \label{fig:compare dif dr}
\end{figure*}

Among the OCs we analyzed, 13 are also included in the study by CG~20 (see the last column of Table~\ref{tab1}). Compared to CG~20, we identified more members in these 13 clusters. 
Generally speaking, we have about twice as many members as CG~20. The CG~20 catalog includes recent improvements to the OC census based on Gaia DR2, previously reported in \cite{cantat2018characterising, cantat2020painting} and \cite{castro2018new, castro2019hunting, castro2020hunting}. 
These works basically use a five-dimensional astrometric based on Gaia DR2 data for clustering, which is usually more general over distance scales in the whole sky region. 
Because they have not been optimized for the membership distribution characteristics of OCs in the nearby sky region, these methods are not applicable in solar neighborhood, and the number of memberships obtained by them is much smaller than that of our method\footnote{To verify this, we repeated our analysis using Gaia DR2 data and observed the same trend.}. 
For many OCs with elongated structures, such as Melotte~111, NGC~2632, it is difficult to observe elongated structures in CG~20. Furthermore, their elongated structures appears not as clearly as our method due to the relatively small number of members. This effect is particularly pronounced for Melotte~111, a close-neighboring OC. 
The members of CG~20 in the literature almost exclusively includes the core of the cluster, whereas our study identifies a population that is twice as large, revealing elongated structures beyond the core. To demonstrate that this effect is not due to the use of different data, we compare the results from our analysis using Gaia DR2 and DR3 with the membership from CG~20 (which uses Gaia DR2) in Fig.~\ref{fig:compare dif dr}. The left and middle panels of Fig.~\ref{fig:compare dif dr} display the distribution of members in sky position and proper motion, respectively. While CG~20 predominantly identifies the compact core region (red points), our study (darkcyan and black markers) captures additional members extending into elongated structures, even when restricted to Gaia DR2 data (darkcyan markers). This consistency indicates that the observed elongated structures are not an artifact of using different data releases. The right panel of Fig.~\ref{fig:compare dif dr} presents a Venn diagram of the membership overlap. This comparison demonstrates that our methodology consistently detects a broader population with more pronounced elongated structures compared to CG~20, irrespective of the Gaia data release used.

\subsection{Elongated structures of OCs}\label{subsec:esoo}

In this section, we discuss the OCs with elongated structures in our cluster member list, and their properties, such as age and morphology. 

\subsubsection{Age Fitting}

\begin{figure*}
    \centering
    \includegraphics[width=0.7\linewidth]{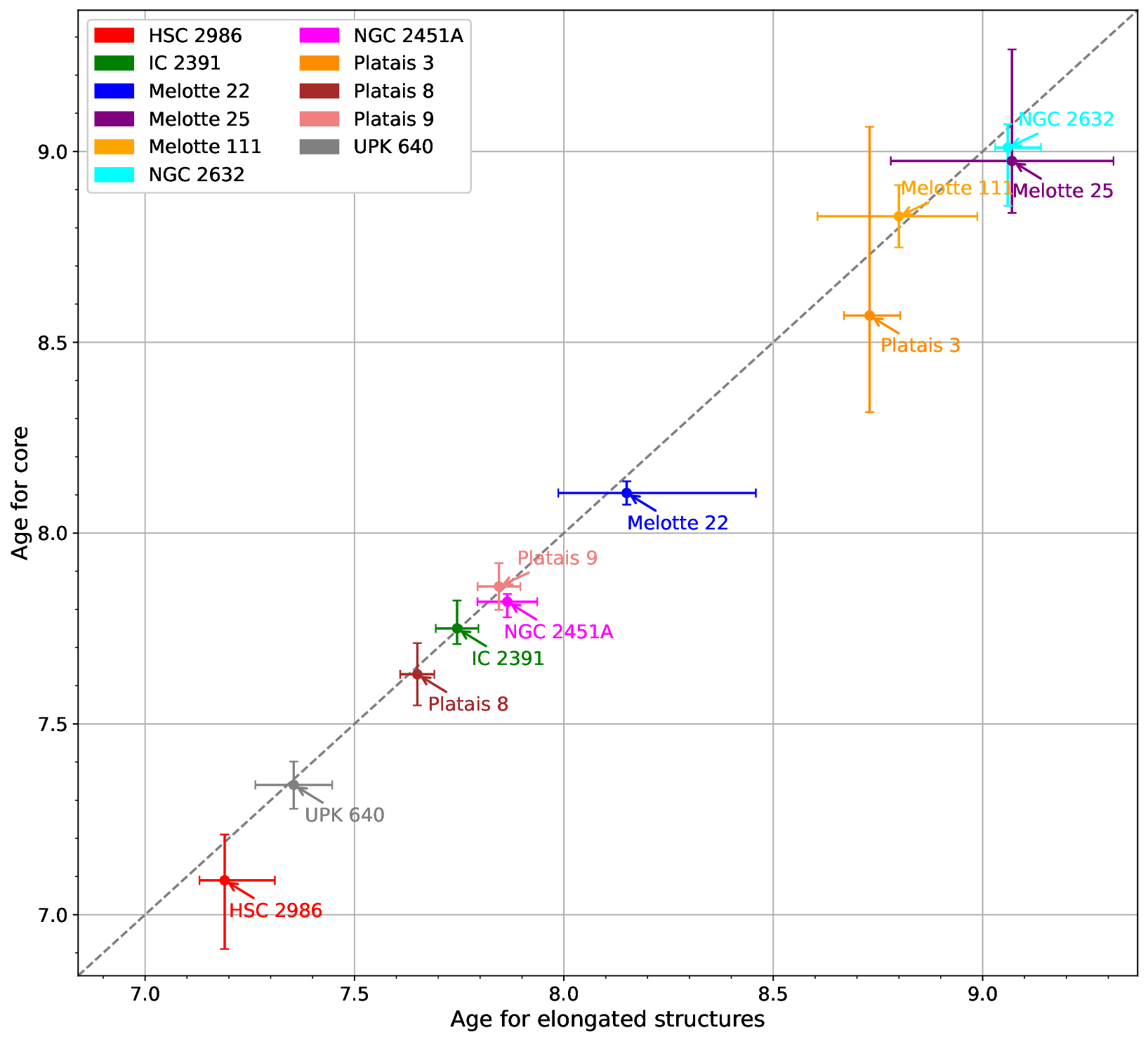}
    \caption{Relationship between the core fitted ages of the OCs HSC~2986, IC~2391, Melotte~22, Melotte~25, Melotte~111, NGC~2632, NGC~2451A, Platais~3, Platais~8, Platais~9, and UPK~640, and the ages fitted for their members beyond the tidal radius. $3\sigma$ error bars are plotted for each OC. The gray dashed line denotes the identity line.
    }
    \label{fig:agecompare}
\end{figure*}

In principle, member stars in an OC are born from the same parental molecular cloud in a single episode of star formation. Most of them are expected to follow a single isochrone in the CMD with the same metallicity and age. 
We employed the PARSEC v1.2s isochrones\footnote{\url{http://stev.oapd.inaf.it/cgi-bin/cmd}} (as presented by \cite{bressan2012parsec}) to fit the ages of OCs in this work. We generated a series of isochrones with ages spanning from $\log{(t/yr})=6.0$ to 9.9, in steps of $\Delta{\log{t}=0.01}$. We adopted the solar metallicity ($Z \simeq 0.0152$~dex) as a fixed parameter in the isochrone models, as studies have shown that variations in metallicity have a negligible impact on determining cluster ages \citep{salaris2004age,cantat2020painting,spina2021galah}. Previous studies have shown that PARSEC isochrones exhibit discrepancies at the low-mass end when compared to observed photometric data of star clusters in CMDs \citep{brandner2023astrophysical}, with the isochrones appearing brighter in absolute magnitude $M_G$ or redder in color (BP$-$RP). This can lead to underestimated ages, particularly for young clusters. To improve the agreement between isochrones and photometric data across the entire mass range of star clusters, we apply the empirical color correction function for PARSEC isochrones proposed by \citet[hereafter Wang~25]{wang2025empirical} to adjust the (BP$-$RP) colors of the selected isochrones.

Before fitting the age, we remove G-band error $\sigma_{G}$ exceeding 0.1~mag and color error $\sigma_{BP-RP}$ exceeding 0.05~mag to ensure high photometric quality. This is an empirical choice based on Wang~25. Our aim is to filter out members with large photometric errors while avoiding overly restrictive criteria. This balance ensures that stars with high photometric uncertainty are excluded without unnecessarily narrowing the sample. To determine the absolute magnitude of each star we utilize the distances from \cite{bailer2021estimating} and based on the extinction data of \cite{capitanio2017three}, we correct the reddening effects caused by the interstellar medium (ISM) absorption of starlight for the members of OCs, following the method used by Wang~25.

During the fitting process, we use the method of \cite{van2023machine} to filter out the stars that are far away from the cluster loci in CMDs. Such an approach allows us to exclude stars that deviate significantly from the cluster loci in the CMDs during the age fitting, such as white dwarfs and some brown dwarfs (typically located at the the lower main sequence with a highly scattered distribution), thereby ensuring a more accurate fit compared to a fitting that includes these stars.

Following the method as described in \cite{liu2019catalog}, we apply the fitting function:
\begin{equation}
\bar{d^ { 2 }}  = \frac { \sum _ { k = 1 } ^ { n } ( \mathbf{x} _ { k } - \mathbf{x} _ { k , n n } ) ^ { 2 } } { n }.
\end{equation}
In this context, $n$ denotes the count of chosen members within a cluster. The variable $\mathbf{x} _ { k }$ indicates the location of the member star in the cluster on the CMD using absolute magnitudes, whereas $\mathbf{x} _ { k , n n }$ refers to the closest point on the isochrone to the member star. We determined the cluster's age by selecting the isochrone with minimal mean CMD discrepancies. Note that we do not get distance modulus and color excess from isochrone fitting. The age uncertainty is obtained by bootstrap resampling on color and absolute magnitude, using 50\% of the data per iteration for 1000 repetitions, with the 16th and 84th percentiles as the uncertainty range. All ages determined using our method are presented in Table~\ref{tab1}.

The age estimates for OCs in this study are compared with those from Wang~25 and Hunt~23, yielding mean differences of 0.025 dex and \textminus 0.269 dex, respectively, and root mean square (RMS) differences of 0.047 dex and 0.327 dex. We also compare the age estimates with those from \cite{dias2021updated}, finding a mean difference of \textminus 0.035 dex and an RMS difference of 0.14 dex. All differences are calculated as literature estimates minus those from this study. Moreover, the ages of those OCs in this work are generally older than those reported in Hunt~23. This is likely because of our use of empirically color-corrected isochrones. These isochrones render the low-mass members of clusters redder, thereby enhancing the match between the model predictions and the observations across the entire mass ranges of the clusters.

\begin{figure*}
    \centering
    \includegraphics[width=1.0\linewidth]{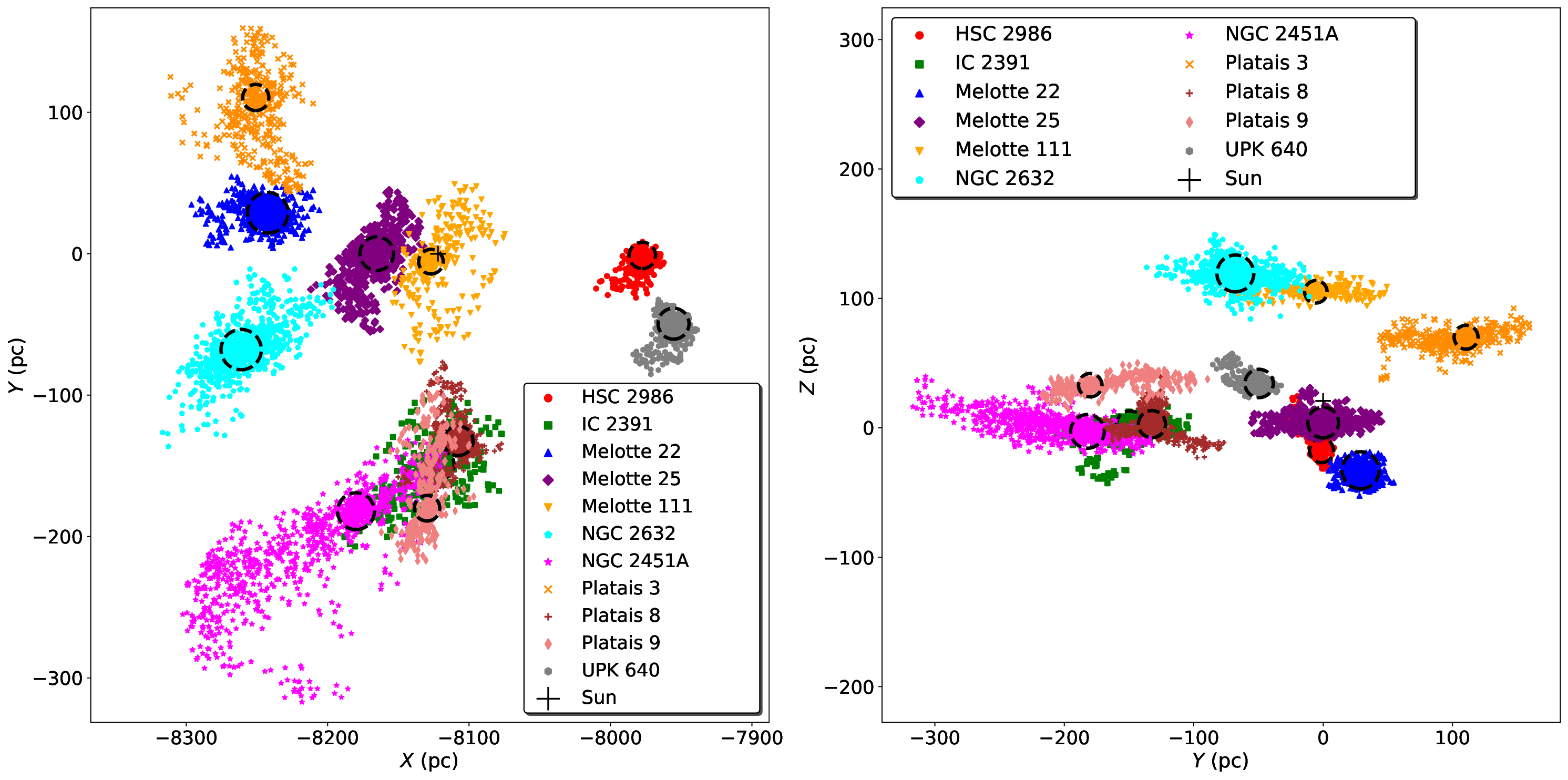}
    \caption{Spatial distribution for HSC~2986, IC~2391, Melotte~22, Melotte~25,
Melotte~111, NGC~2632, NGC~2451A, Platais~3, Platais~8, Platais~9, and UPK~640 in Galactocentric Cartesian coordinates. The black circle denotes the tidal radius of each OC (typical errors are 0.07–0.13 pc). The black cross symbol indicates the position of the Sun. The morphology and age of these OCs are discussed in Sec.~\ref{subsec:esoo}.}
    \label{fig:elongated}
\end{figure*}

\subsubsection{Elongated structures}

We selected OCs with elongated structures, dividing each OC into two components: one within (core) and the other outside (elongated structures) the tidal radius, and fitted the age for each component separately. We define elongated structures as relatively dense cores accompanied by stretches of structures aligned with the Galactic plane \citep{2004AJ....128.2306C,2021ApJ...912..162P}. To identify such structures, we first examined the radial profile distribution to confirm a declining trend, ensuring the presence of a dense core. Next, we analyzed the angular distribution, which refers to the cluster's star distribution by angle, using the Galactic coordinates in the X-Y plane as the reference plane, starting from the direction of the Galactic center and rotating counterclockwise. Note that we required at least one of 10 equal angular bins to contain more than 15\% of the total stars, a threshold indicating elongation beyond isotropic expectations (10\%). Additionally, the maximum radius in the elongated direction had to be at least 1.5 times the tidal radius. These criteria were supplemented by visual inspection to validate the results. 
The tidal radius was calculated using the method in \cite{2021ApJ...912..162P}, in which the masses of individual members, derived from the nearest point in the best-fit isochrone, sum up to obtain the total mass of the cluster. 

As shown in Fig.~\ref{fig:agecompare}, for each of the elongated OC we selected based on the previously mentioned criteria, the fitted ages of the core (within the tidal radius) and the portions beyond the tidal radius are in agreement within a $3\sigma$ level. 
This suggests that members detected far from the core likely originate from the same molecular clouds as the core members. The age distribution of these OCs is broad and can be divided into three groups: those older than 8.5~dex (about 316~Myr), those between 7.5~dex and 8.5~dex (roughly 31~Myr to 316~Myr), and those younger than 7.5~dex.

\begin{itemize}
    \item[(i)]  Age older than 8.5~dex (about 316~Myr):\\   
    The OCs in the oldest group are Melotte~25, NGC~2632, and Melotte~111 and Platais~3, 
    including a leading part toward the galactic center and a trailing part (see Fig.~\ref{fig:elongated}). 
    They are old enough for the tidal field to shape their morphology as they rotate around the galaxy, resulting in the formation of tidal tails, this is consistent with other work \citep{zhang2020diagnosing, jerabkova2021800, ye2021extended}. 
    
    \item[(ii)]  Age between 7.5~dex and 8.5~dex:\\
For OCs in the intermediate age group, we do not find a clear pattern in their morphology. Some of them own a long arm structure (NGC~2451A, see Fig.~\ref{fig:galacompare} and Fig.~\ref{fig:morphology}), but 
some owns sparse coronae (Melotte~22, IC~2391) or irregular dissolving shapes (Platais~8, Platais~9). The OCs (IC~2391, NGC~2451A, and Platais~8) spatially overlap to some extent, and these groups are sparsely connected in velocity space. This suggests that these three OCs may have a common origin, and similar results were seen within \cite{2019A&A...624L..11F} and \cite{meingast2021extended}. 

    \item[(iii)]  Age younger than 7.5~dex:\\
The two youngest OCs, HSC~2986 and UPK~640, have similar ages and are located in the Scorpius-Centaurus (Sco-Cen) constellation, known for its star-forming regions \citep{ratzenbock2023significance}. Both exhibit a dense core with a filamentous elongated structures. Their origin may be linked to the star formation history in Sco-Cen, where population ages vary with distance \citep{kerr2021stars}. HSC~2986 and UPK~640 are younger and may have formed from turbulent processes triggered by earlier generations of star formation. This requires further study of their formation.
\end{itemize}

As shown in Fig.~\ref{fig:elongated}, OCs of the same age group tend to be close together, likely due to their formation from a giant molecular cloud through sequential star formation caused by density fluctuations \citep{rizzuto2011multidimensional, jerabkova2019stellar, krause2020physics}. All OCs discussed here extend parallel to the Galactic plane, forming tidal tail-like structures. This morphology is consistent across different ages. In older clusters, it is mainly due to tidal field effects. 
In younger clusters, while the tidal field contributes to their shape, their morphology is more influenced by initial formation conditions, gas expulsion during star formation, and the molecular cloud's shape in the formation region \citep{2010ApJ...723..492R, 2012MNRAS.427.1940P}.

In \cite{2021ApJ...912..162P}, it was found that tidal tails of OCs align with the Galactic plane, as these structures form when the cluster is stretched by the Galactic tide. However, this alignment does not apply to all elongated structures, particularly filamentary (young) OCs, which are not necessarily affected by the tidal force and may have extended directions that are either parallel to or inclined with respect to the Galactic plane. Our definition of elongated structures is unable to recognize OCs that are elongated in directions nearly perpendicular to the Galactic plane. We performed ellipsoid fitting\footnote{\url{https://github.com/marksemple/pyEllipsoid_Fit}} on all the OCs in this study, calculating the angle between their semimajor axis and the Galactic plane. OCs older than 8.5 dex align closely with the Galactic plane, while younger clusters (younger than 8.5 dex) exhibit a wider range of orientations, with a median inclination angle of $\approx 17.6^{\circ}$, extending from nearly parallel to steeply inclined, with angles reaching up to $\approx 69.7^{\circ}$. HSC~2505 and OCSN~96 are younger clusters with an inclination angle greater than $50^{\circ}$ (2 out of 23), for which our definition of elongated structures might not be fully applicable.

\subsubsection{The oldest group: dissolving Open Clusters}\label{subsec:doc}

\begin{figure*}
    \centering
    \includegraphics[width=1.\linewidth]{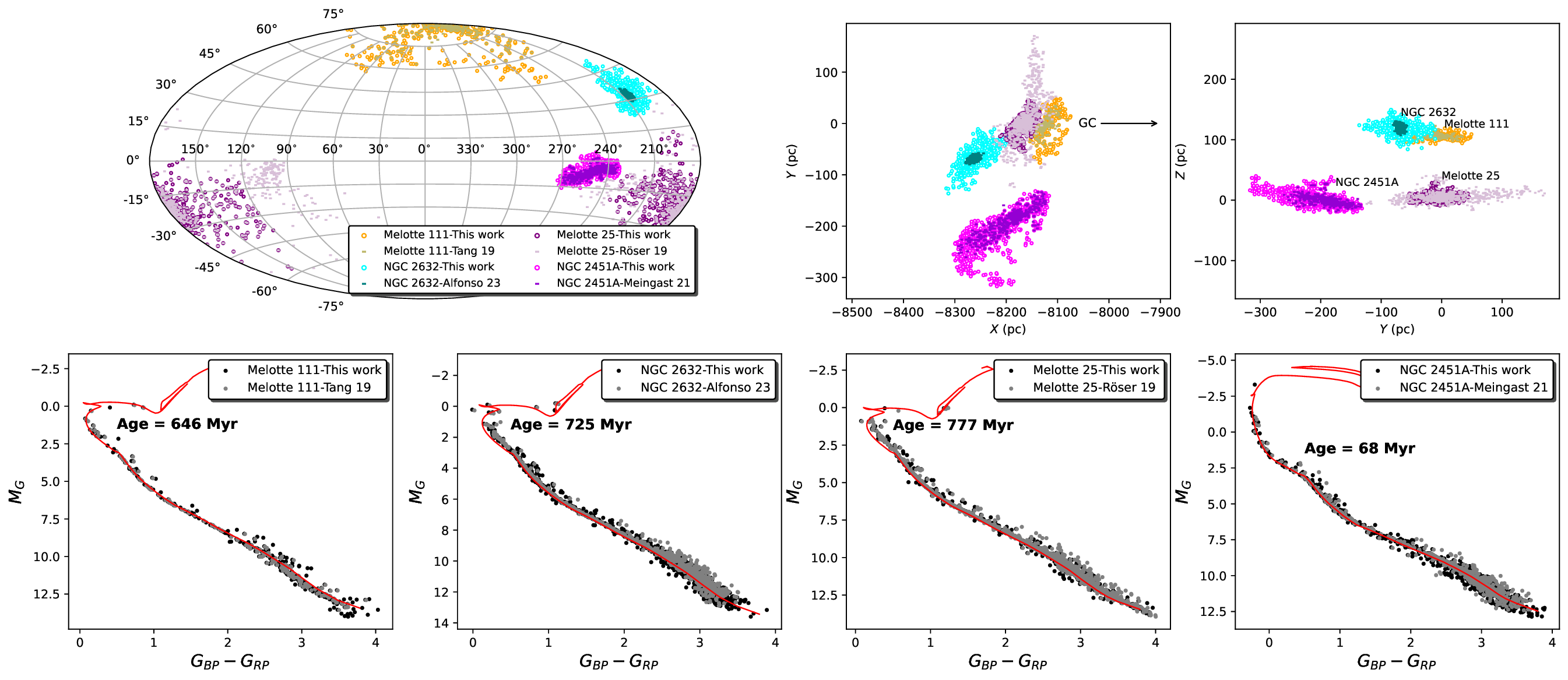}
    \caption{Distribution of members of Melotte~111, Melotte~25, NGC~2632, NGC~2451A in this work and in \cite{tang2019discovery} (Tang~19), \cite{roser2019hyades} (Röser~19), \cite{alfonso2023gaia} (Alfonso~23), \cite{meingast2021extended} (Meingast~21) respectively. The upper left panel shows the distribution of the selected clusters in Galactic coordinates, while 2 panels on the upper right show their spatial distribution in Galactocentric Cartesian coordinates. Directions to galactic centers are given by black arrows. The four plots in the lower panel show the CMDs for individual OCs, where the red solid curves represent the empirically corrected PARSEC isochrones in \cite{wang2025empirical} with a solar metallicity, labeled with their age.}
    \label{fig:galacompare}
\end{figure*}

OCs in the oldest group exhibit tidal structures, typically identified through various methods in other studies. However, our methodology enabled us to identify all these structures and compile a more comprehensive membership list than previous studies. As aforementioned in Sec.~\ref{subsec:moe}, our method can detect elongated structures of open clusters. Compared to Hunt~23, our method generally identifies more members in these structures. Some structures result from open clusters losing stars due to the Milky Way gravitational tidal forces, reflecting the process of cluster evolution and dissipation. As clusters move along their orbits, different gravitational forces pull stars out of the cluster boundaries, forming leading and trailing tidal tails. 
This feature is probed in Melotte~111 (Coma~Berenices). 

Fig.~\ref{fig:galacompare} illustrates the distribution of members of Melotte~111, Melotte~25, NGC~2632, and NGC~2451A as identified in this study, in comparisons with Tang~19, R{\"o}ser~19, Alfonso~23, and Meingast~21. The upper-left panel shows the clusters distribution in Galactic coordinates, while the two upper-right panels depict their spatial distribution in Galactocentric Cartesian coordinates (The Sun located at $(X,Y,Z) = (-8122,0,20.8)$~pc \citep{reid2004proper}). The lower panels present the color-magnitude diagrams (absolute magnitude) for each OC, with red curves showing best-fit empirically corrected PARSEC isochrones in Wang~25, labeled with age (assuming solar metallicity). 

As shown in Fig.~\ref{fig:galacompare}, we clearly find leading and trailing tidal tails of Melotte~111 due to the Milky Way differential gravitational force, 
in agreement with Tang~19. 
\cite{2021ApJ...912..162P} also conducted membership validation for Melotte 111 and analyzed its morphology. Compared to their work, we identified more members and observed more distinct elongated structures, with the elongation direction being about 20 pc longer than in their study. This trend also appears in other younger clusters that overlap with both our analysis and theirs, such as NGC~2451A and IC~2391. For NGC~2451A, our membership list is nearly four times larger than theirs, and we identified prominent leading and trailing structures that were not detected in their study. Specifically, the leading structures extend approximately 50 pc, while the trailing structures span around 100 pc, with a vertical distribution of about 40 pc above and below the Galactic plane. Most of their identified members are located within the tidal radius of the cluster, whereas our results reveal a far more extensive spatial distribution. Similarly, for IC 2391, our membership is approximately three times larger, and nearly all of their identified members are located within the tidal radius, in stark contrast to the broader structures we detected.

\begin{figure*}
    \centering
    \includegraphics[width=0.7\linewidth]{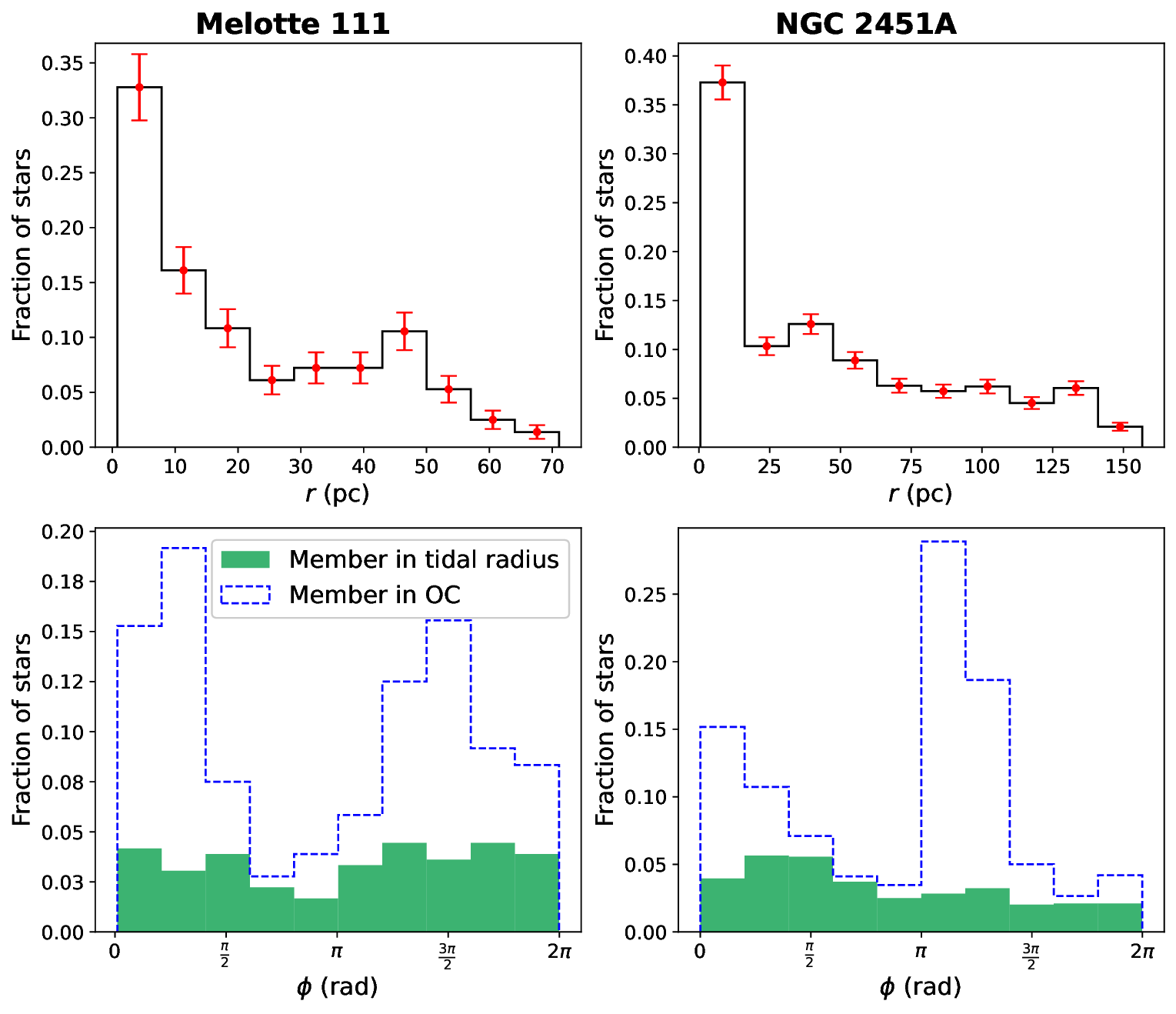}
    \caption{Radial and angular distribution of selected OCs Melotte~111, NGC~2451A in this paper. The top panels show histograms of the fraction of stars at different radii from the cluster center, with error bars plotted on each bin. The lower panels show the angular distribution of the fraction of members at the tidal and maximum radii of the cluster whose distance from the cluster center is less than that radius, using the Galactic coordinates in the X-Y plane as the expansion plane, starting from the direction of the center of the Milky Way and rotating counterclockwise. 
    For these OCs, two distinct peaks can be seen in their lower plots, indicating their leading and trailing arms.
    }
    \label{fig:morphology}
\end{figure*}

Figure~\ref{fig:morphology} shows the radial and angular distributions of members in Melotte~111 (an old cluster) and NGC~2451A (a young cluster) illustrating the morphological structures of clusters at different ages. The top panels show histograms of the fraction of stars at different radii from the cluster center, providing insight into their radial distribution. The lower panels illustrate the angular distribution of members within the tidal and maximum radii, plotted in the Galactic X-Y plane, starting from the direction of the Galactic center and rotating counterclockwise. Two prominent peaks can be seen for these two OCs, corresponding to their leading and trailing arms. Although both exhibit two distinct peaks, the reasons for OCs of different ages to show this characteristic are different, as we mentioned in the last paragraph of Sec.~\ref{subsec:esoo}. This features are seen in OCs, such as NGC~2632 and Melotte~25 (Hyades), in Fig.~\ref{fig:galacompare}. For Melotte~25, we find leading and trailing tidal tails similar to R{\"o}ser~19, whereas Hunt~23 does not detect leading tails. We identify nearly as many members as R{\"o}ser~19, but their tails are more extended due to using the convergence point method instead of a clustering algorithm. NGC~2451A is not in the oldest group, but it has similar structures to the three OCs in the oldest group, therefore it is included in Figs.~\ref{fig:galacompare} and \ref{fig:morphology} for comparison.

\subsubsection{The group younger than 8.5~dex: elongated Open Clusters}\label{subsubsec:nptt}

\begin{figure*}
    \centering
    \includegraphics[width=0.8\linewidth]{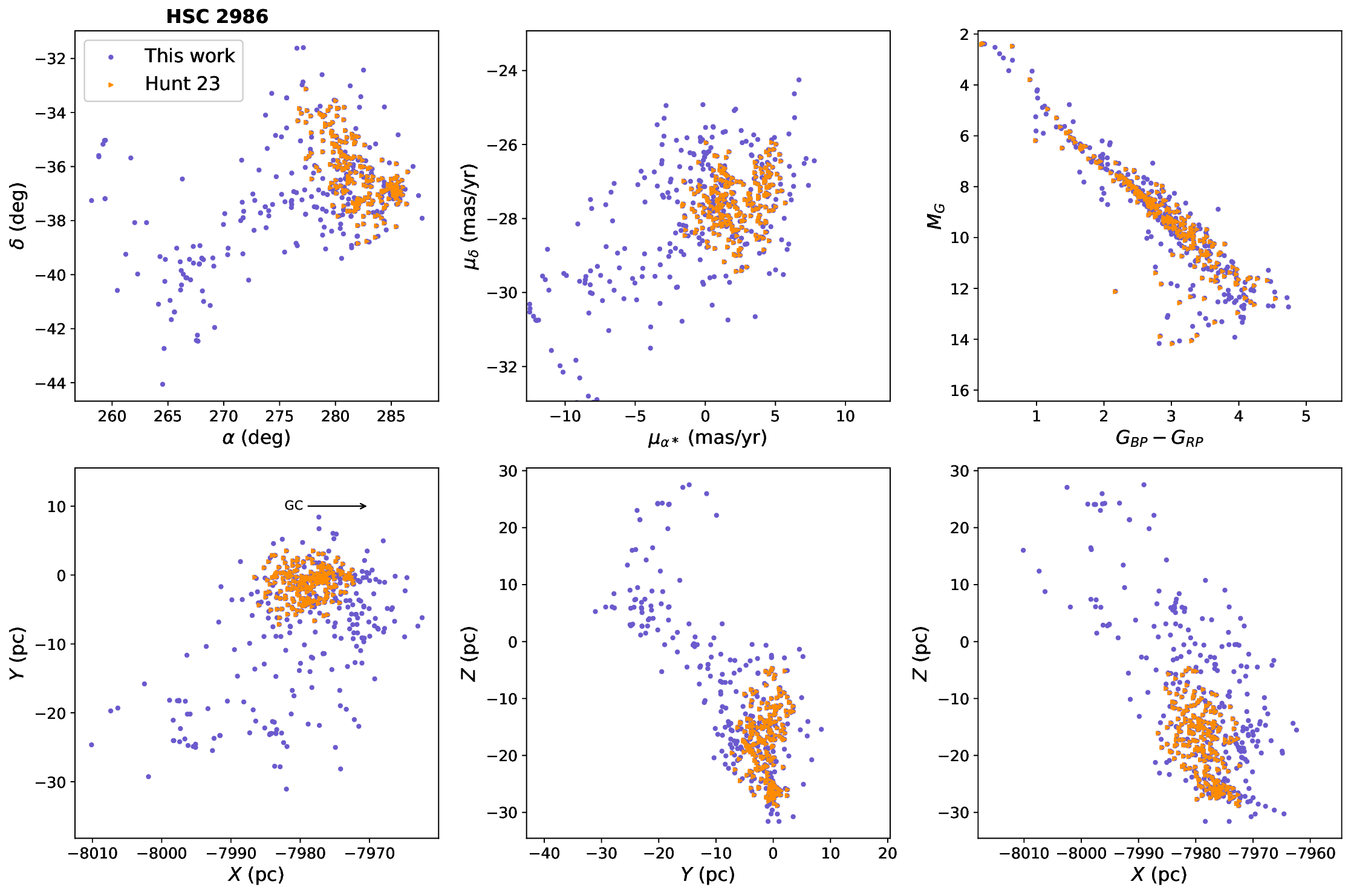}
    \caption{Distribution of the HSC~2986 of our members and the members of Hunt~23 on the position of the sky, the \texttt{proper motion}, the CMD and the Galactocentric Cartesian coordinates.}
    \label{fig:hsc2986distribution}
\end{figure*}

\begin{figure*}
    \centering
    \includegraphics[width=1.\linewidth]{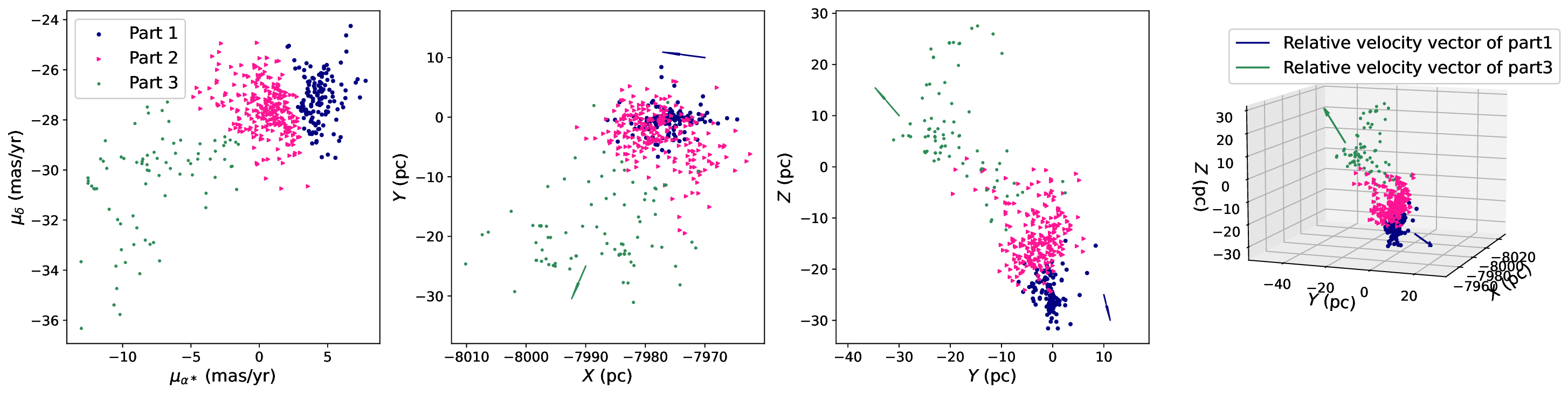}
    \caption{\texttt{Proper motion} and spatial distribution for candidate members of HSC~2986 in this work. Its distribution on the \texttt{proper motion} is divided into three parts, Part 1, Part 2 and Part 3, which correspond to the near-core members, the core members and the possible extended structures, respectively. The colored arrows in the middle two plots indicate the relative mean projected velocity vector of the corresponding part with respect to the mean projected velocity vector of part 2, whereas the colored arrows in the right plot indicate the relative velocity vectors of the corresponding part with respect to the mean velocity vector of part 2, $(U,V,W) = (6.9, 228.2, 0.7) \textnormal{ km s}^{-1}$.}
    \label{fig:hsc2986morphology}
\end{figure*}

\textbf{\underline{Ages between 7.5~dex and 8.5~dex}:}

The morphology of younger OCs is more complicate than that of the oldest groups. 
For instance, Melotte~22 (Pleiades) has been linked to nearby moving groups such as the AB Doradus group \citep{2021ApJ...915L..29G}. While Melotte~22 may possess a leading structures, both our method and the one in Meingast 21 do not detect it, likely because the leading part lies behind the Pleiades, making it challenging for Gaia to distinguish. Despite our membership list being approximately 600 members more than that in Meingast 21, the additional members are mostly concentrated in the core and trailing structures. 
Similarly, in other OCs we analyzed that were also studied by Meingast 21 (NGC~2451A, IC~2391, Platais~9), our membership lists typically contain more members than theirs, and while the extended structures we detect are generally consistent with their results, our identified structures are more extensive. For example, in NGC~2451A (see Fig.~\ref{fig:galacompare}), our membership list contains about twice as many members as theirs, with the structures extending in nearly the same direction. While the leading structures in our results are comparable in length to those identified by Meingast 21, the trailing structures are about 50 pc longer and show an additional extension of about 20 pc vertically above the galactic plane compared to Meingast 21's results. 

Similarly, young OCs such as UPK~640 and HSC~2986 exhibit trailing structures but lack visible leading counterparts. For UPK~640, our findings closely match a trailing elongated structures found by Hunt~23. However, a newly discovered young cluster (HSC~2986) has a sparse trailing morphology in our result, while Hunt~23 has identified only its core. These two clusters are too young for tidal forces to form tidal tails, thus their elongated morphologies are not tidal tails \citep{dinnbier2020tidali, dinnbier2020tidalii, 2021ApJ...912..162P}.

Platais~8, first identified in \cite{platais1998search}, exhibits an elongated morphology with a denser core and more extensive leading structures than IC~2391 in this study, suggesting ongoing dissolution. While \cite{2018A&A...615A..12Y}, using Gaia DR1 \citep{brown2016gaia}, HSOY \citep{altmann2017hot}, and HIPPARCOS \citep{van2007validation}, have identified only 38 members near the core, they miss the extended structures revealed in this study. According to \cite{2021ApJ...915L..29G}, Theia~92 \citep[identified by][]{kounkel2019untangling} includes the core of Platais~8 and a possible trailing tidal tail behind this core. Our analysis confirms that Platais~8's trailing tidal tail overlaps significantly with Theia~92. Additionally, applying algorithm in \cite{gagne2018banyan}, we identified partial overlap between the leading structure and Carina, Columba, and Theia~113, linking these groups to Platais~8 core. 

However, the fact that the leading structure is at a greater distance from the Galactic Center than the core suggests that the cluster's dissolution is unlikely to be driven by tidal forces alone, which indicates the need for further study of the cluster’s dissolution mechanisms in the future.

\textbf{\underline{HSC~2986, a young elongated Open Cluster}:}
 
For the first time, we report evidence of an elongated morphology in HSC~2986, a young cluster identified by Hunt~23 at a distance of approximately 150~pc. Our analysis reveals sparse, extended structures in \texttt{sky position}, \texttt{proper motion}, and spatial distribution, with roughly twice the membership reported by Hunt~23, as illustrated in Fig.~\ref{fig:hsc2986distribution}. The upper panels of Fig.~\ref{fig:hsc2986distribution} show the position on the sky, the \texttt{proper motion}, and the CMD for the HSC~2986 members identified in this study compared to those from Hunt~23. The lower panels show the distribution of the members in Galactocentric Cartesian coordinates. Our results display a sparse and extended structure that was not fully captured in Hunt~23, demonstrating the broader spatial extent of the cluster members identified by our method.

In Fig.~\ref{fig:hsc2986morphology}, we present the \texttt{proper motion} and spatial distribution of candidate members of HSC~2986 identified in this work, categorized into three parts: near-core members (Part~1), core members (Part~2), and possible extended structures (Part~3). The figure highlights the relative mean projected velocity vectors of each part, with colored arrows in the middle panels showing their motion relative to the core members (Part~2), and arrows in the right panel depicting their relative velocity vectors with respect to the mean velocity vector of Part~2. The \texttt{proper motion} distribution separates the cluster into three distinct groups, divided based on observed patterns rather than strict numerical criteria, as shown in the left panel of Fig.~\ref{fig:hsc2986morphology}. Part~1 (blue) corresponds to the right side of the distribution, while Part~2 and Part~3 (pink and green) represent the dense central area and the sparse extended outskirts, respectively. Spatially, Part~1 and Part~2 are closely positioned yet distinctly separated, with Part~3 forming an extended structure approximately 30~pc in length. Taking the mean velocity of Part~2, $(U_2, V_2, W_2) = (6.9, 228.2, 0.7) \, \textnormal{km s}^{-1}$, as a reference, the relative velocity vectors of Part~1 and Part~3 are $(U_{1,2}, V_{1,2}, W_{1,2}) = (-2.1, 0.3, -1.1) \, \textnormal{km s}^{-1}$ and $(U_{3,2}, V_{3,2}, W_{3,2}) = (-0.6, -1.4, 1.6) \, \textnormal{km s}^{-1}$, respectively.

The middle panels of Fig.~\ref{fig:hsc2986morphology} depict the projections of the relative velocity vectors in Galactocentric Cartesian coordinates. These reveal that Part~3 is moving away from the Galactic center, opposite to the Galactic rotation, and upward perpendicular to the Galactic plane, while Part~1 also moves away from the Galactic center but in directions distinct from Part~3. This trend is further emphasized in the right panel, a three-dimensional projection map of HSC~2986 spatial distribution. 
Part~3 moves away from Part~2 at a speed of approximately $2.2~\textnormal{km s}^{-1}$, with its center located about 20~pc from the edge of Part~2. Assuming that the relative velocity between Part 3 and Part 2 was consistent throughout, the center of Part~3 would have reached the boundary of Part~2 around 9 million years ago. This aligns with the estimated age of HSC~2986, suggesting that Part~3 likely originated from the same molecular cloud as the cluster core. We used the {\sc galpy} package \citep{bovy2015galpy} to integrate the orbits of member stars with \texttt{RV}, tracing them backward in time to determine their birth locations. For the gravitational potential of the Milky Way, we assumed the {\sc MWpotential2014} model, with the local standard of rest (LSR) velocity set to $V_{\text{LSR}} = 229~\textnormal{km s}^{-1}$ \citep{eilers2019circular}. This further confirms that center of Part~3 would have reached the boundary of Part~2 around 9 million years ago. However, when tracing back to the age of the cluster (approximately 13 Myr), their relative position show little variation compared to those at 9 Myr, suggesting that the cluster likely had a loose, filamentary morphology at the time of its formation.

\section{Summary}\label{sec:con}

In this study, we applied a refined membership determination method, utilizing the precise astrometric data from Gaia DR3, to identify members of 30 open clusters within 200~pc of the Sun. By correcting for projection effects that distort the apparent motions of stars, our approach enables a more comprehensive assessment of cluster memberships and structural analysis of these stellar groups. This work offers valuable insights into the internal dynamics of open clusters, crucial for understanding their evolution and interactions with the Galactic environment. The key findings are summarized as follows:

\begin{enumerate}

\item For most of the selected open clusters (OCs), our method identifies significantly more members compared to previous studies, with an average membership approximately 1.5 times that of Hunt~23, 2.4 times that of Tar~22, and 2.6 times that of CG~20.

\item Using the identified members, we have determined the ages of all selected OCs. 
For clusters with elongated morphologies, we further divide members based on tidal radii and fit their ages separately, confirming that the members are coeval.

\item Our method effectively detects the elongated morphology of nearby open clusters, while typically identifying more cluster members than previous work. For OCs previously observed to exhibit a dissolving trend (e.g., Melotte~22, Platais~8), our results agree with their reported expansion trends but include more members and reveal clearer structural features, highlighting the efficiency of our approach.

\item We identified HSC~2986 as having an elongated structures. Spatial and kinematic analyses confirm and support its members are coeval.

\end{enumerate}

Future studies should continue to refine the current methodologies, expand the sample size, and integrate multiwavelength data to enhance our understanding of open cluster evolution. Removal of contamination from data sets will be crucial to improving the accuracy of these studies. Additional efforts to correlate observed structural features with theoretical models of cluster dynamics would also be beneficial in confirming the trends suggested by this study and further elucidating the complex processes that govern the life cycles of open clusters.

\section*{Acknowledgements}

We sincerely thank Dr.~Hunt for the kind help. We thank the anonymous reviewer for the valuable comments, which have significantly enhanced the clarity of this paper. We also thank Dr.~Yin Li for the assistance provided during the preparation of this paper. We are grateful to Dr.~Lu Li for the detailed and enlightening discussion. This work is supported by the National Key Research and Development Program of China (No. 2023YFA1608100). YST are supported by the National Key Research and Development Program of China (No. 2022YFF0503304), and the Project for Young Scientists in Basic Research of the Chinese Academy of Sciences (No. YSBR-092). Xiaoying Pang acknowledges the financial support of the National Natural Science Foundation of China through grants 12173029 and 12233013. X.F. thanks the support of the National Natural Science Foundation of China (NSFC) No. 12203100 and the China Manned Space Project with No. CMS-CSST-2021-A08.
This work made use of data from the European Space Agency (ESA) mission Gaia (\url{https://www.cosmos.esa.int/gaia}), processed by the Gaia Data Processing and Analysis Consortium (DPAC, \url{https://www.cosmos.esa.int/web/gaia/dpac/consortium}). Funding for the DPAC has been provided by national institutions, in particular the institutions participating in the Gaia Multilateral Agreement. This work also made use of the SIMBAD database and the VizieR catalog access tool, both operated at CDS, Strasbourg, France.

\software{{\sc Astropy}~\citep{astropy:2013, astropy:2018, astropy:2022},{\sc galpy}~\citep{bovy2015galpy}}

\bibliography{main}
\bibliographystyle{aasjournal}

\begin{appendix}
\section{Distribution of members in OCs}\label{app:doo}

In Fig~\ref{fig:apoc}, we present the distributions of member sky positions, proper motions, and CMDs for all OCs in this work and compare them with the results from Hunt~23.

\renewcommand{\thefigure}{A.\arabic{figure}}
\setcounter{figure}{0}
\begin{figure*}[b]
\centering
    {%
    \includegraphics[width=0.9\textwidth, trim=0.1cm 0.9cm 0.0cm 1.6cm]{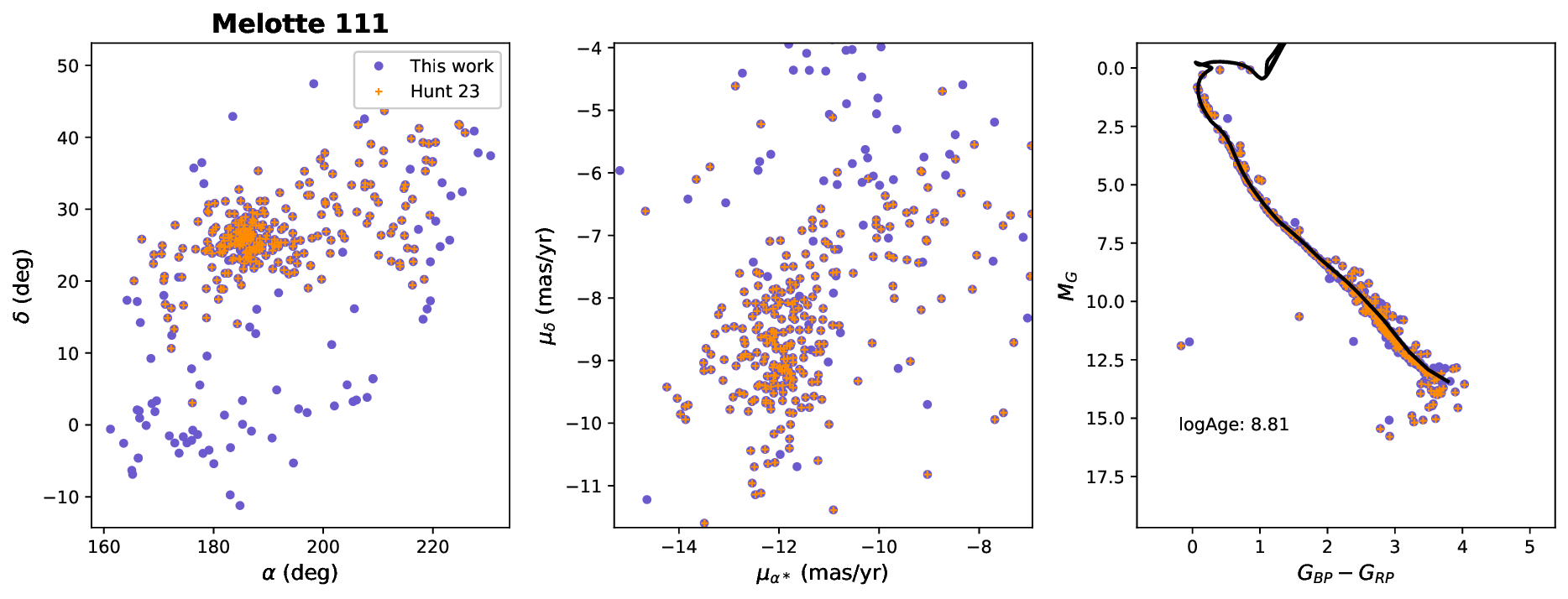}
    }
    \\[1.1cm]
    {
    \includegraphics[width=0.9\textwidth, trim=0.1cm 0.9cm 0.0cm 1.6cm]{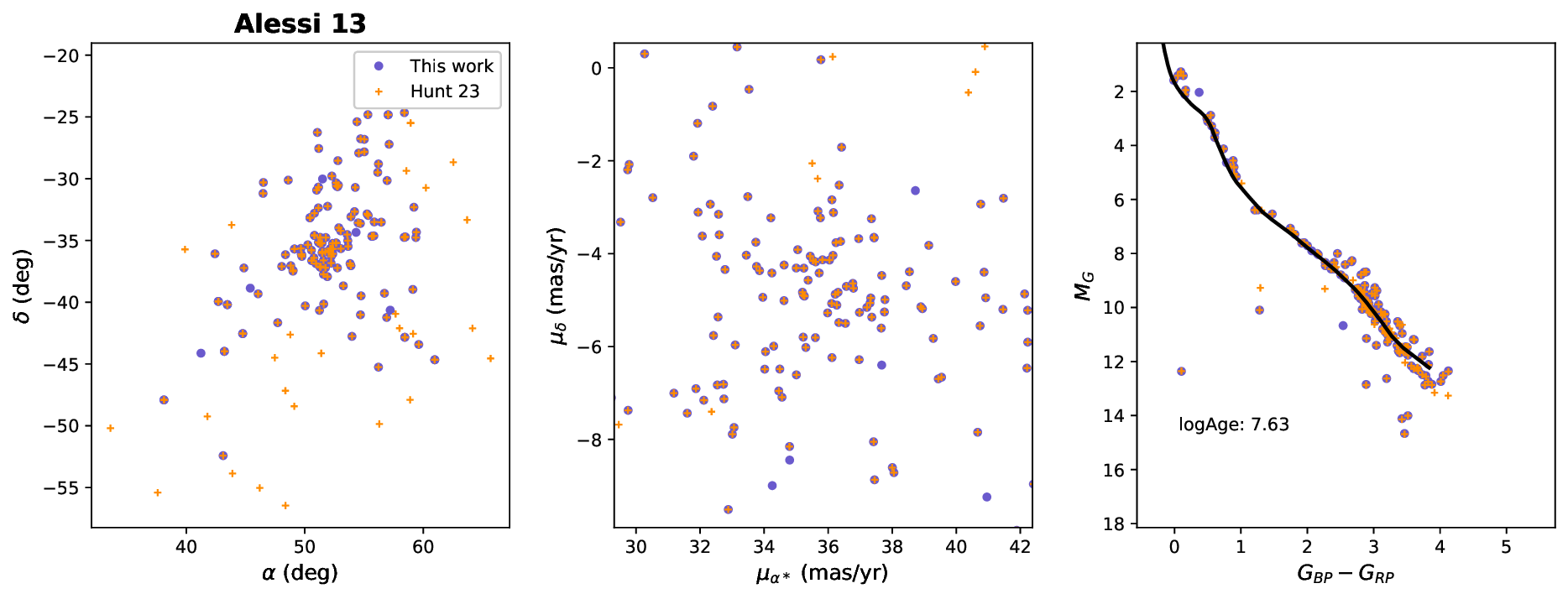}
    }
    \\[1.1cm]
{%
\includegraphics[width=0.9\textwidth, trim=0.1cm 0.9cm 0.0cm 1.6cm]{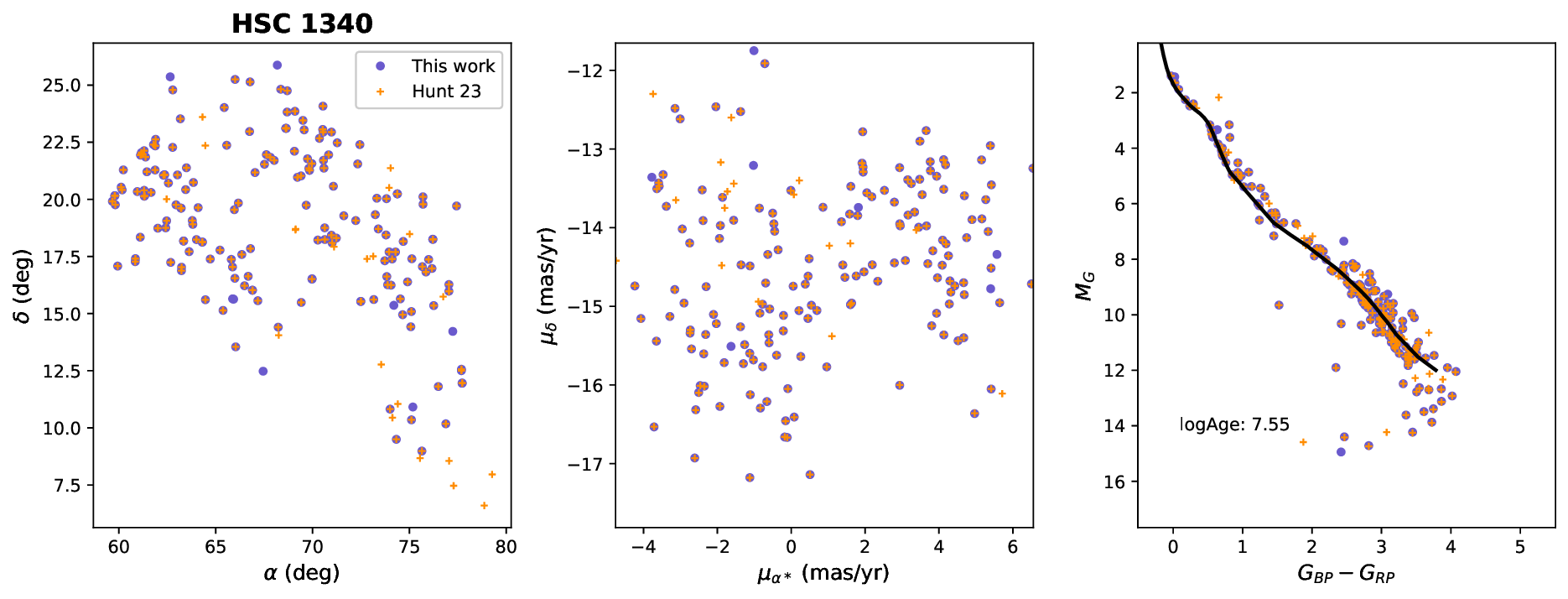}
}
\\[0.2cm]

\caption{Distribution of our members (solid blue circles) and the members of Hunt~23 (orange crosses) on \texttt{sky position}, \texttt{proper motion} and CMD for OCs in this study, the black solid line curves represent the best-fit empirically corrected PARSEC isochrones in Wang 25 (see Sec.~\ref{subsec:esoo} for details).}
\label{fig:apoc}
\end{figure*}

\setcounter{figure}{0}
\begin{figure*}
\centering
{%
\includegraphics[width=0.9\textwidth, trim=0.1cm 0.9cm 0.0cm 1.6cm]{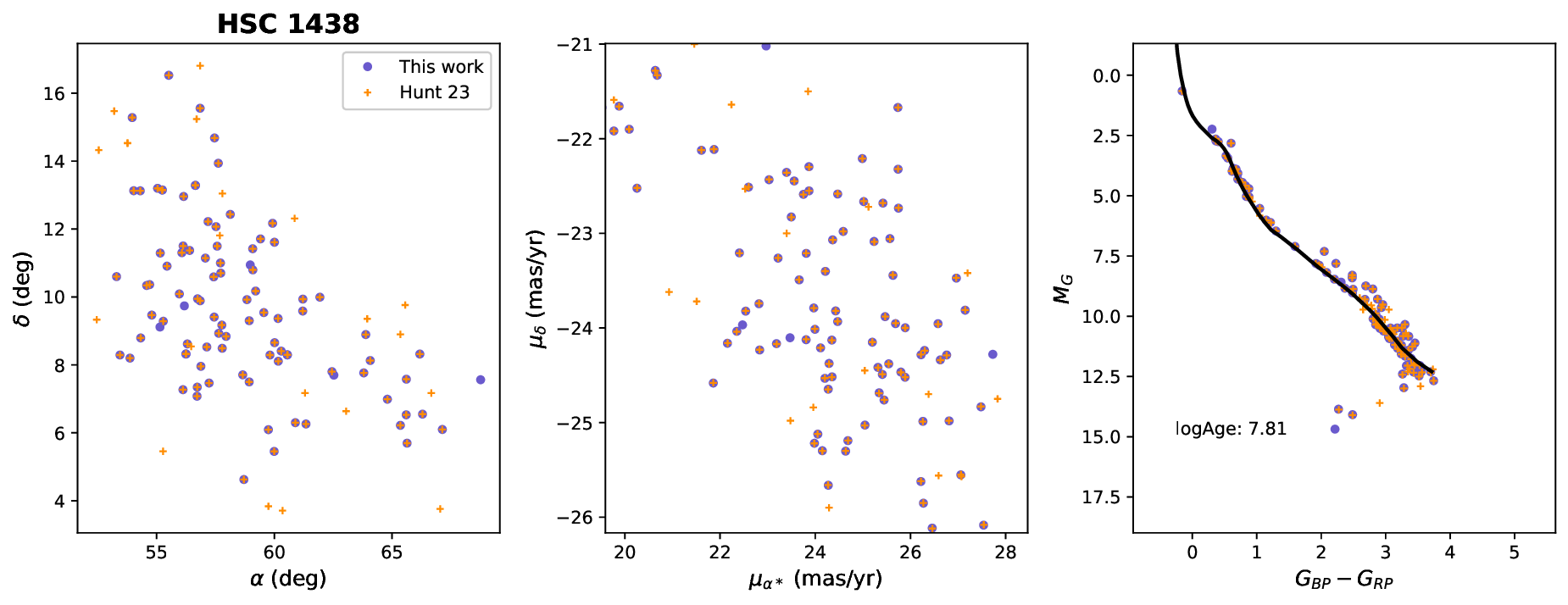}
}
\\[1.1cm]
{%
\includegraphics[width=0.9\textwidth, trim=0.1cm 0.9cm 0.0cm 1.6cm]{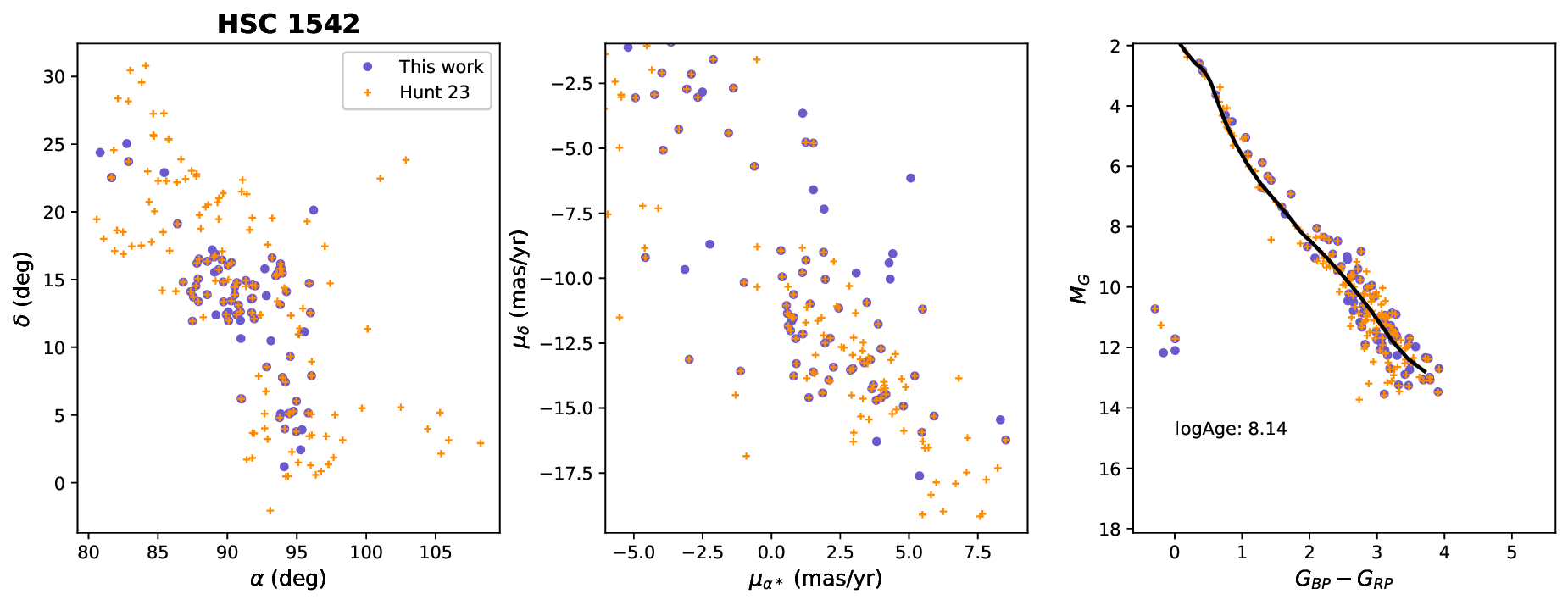}
}
\\[1.1cm]
{%
\includegraphics[width=0.9\textwidth, trim=0.1cm 0.9cm 0.0cm 1.6cm]{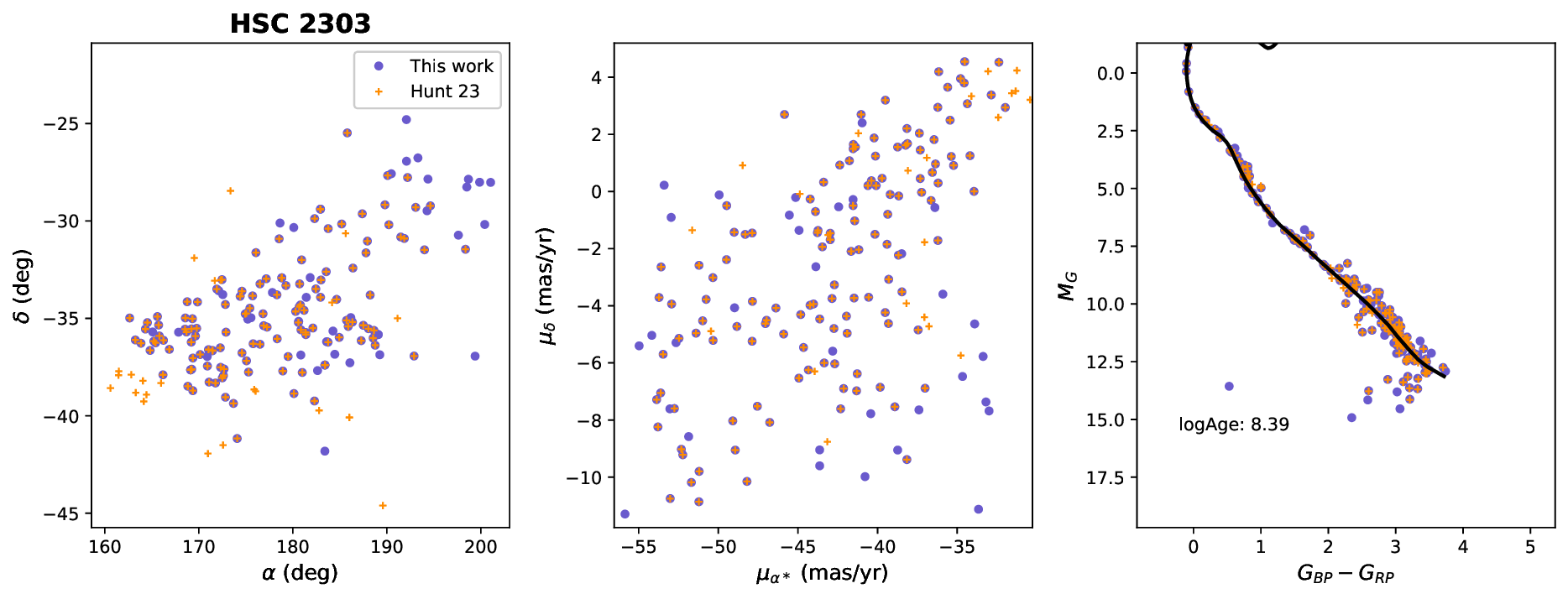}
}
\\[1.1cm]
{%
\includegraphics[width=0.9\textwidth, trim=0.1cm 0.9cm 0.0cm 1.6cm]{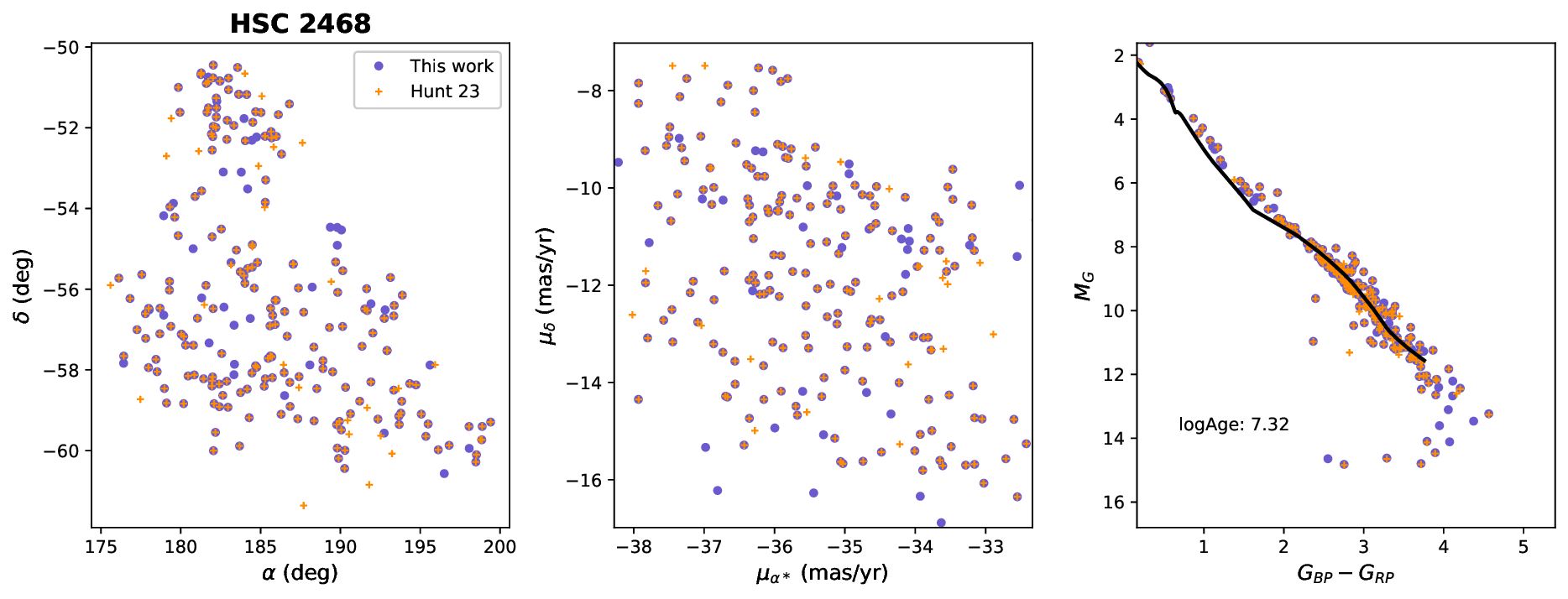}
}
\\[0.2cm]
\caption{Continued Fig~\ref{fig:apoc}}
\label{fig:apoc}
\end{figure*}

\setcounter{figure}{0}
\begin{figure*}
\centering
{%
\includegraphics[width=0.9\textwidth, trim=0.1cm 0.9cm 0.0cm 1.6cm]{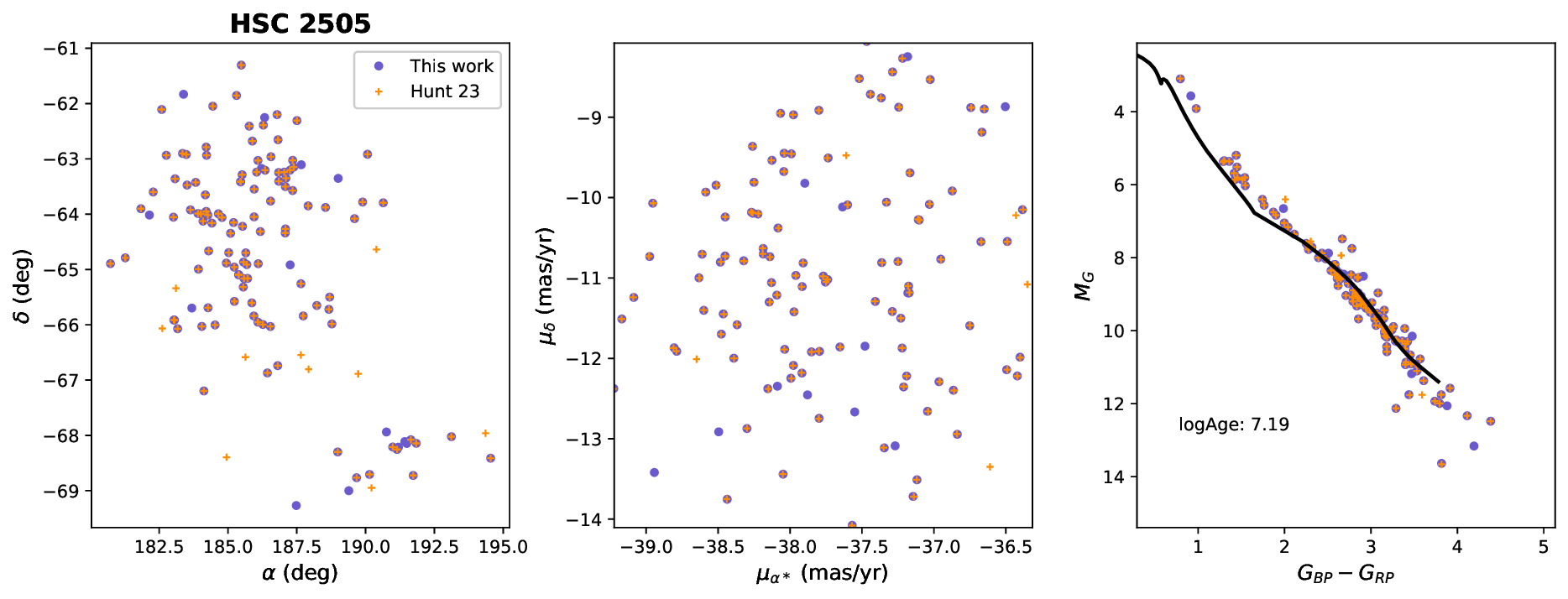}
}
\\[1.1cm]
{%
\includegraphics[width=0.9\textwidth, trim=0.1cm 0.9cm 0.0cm 1.6cm]{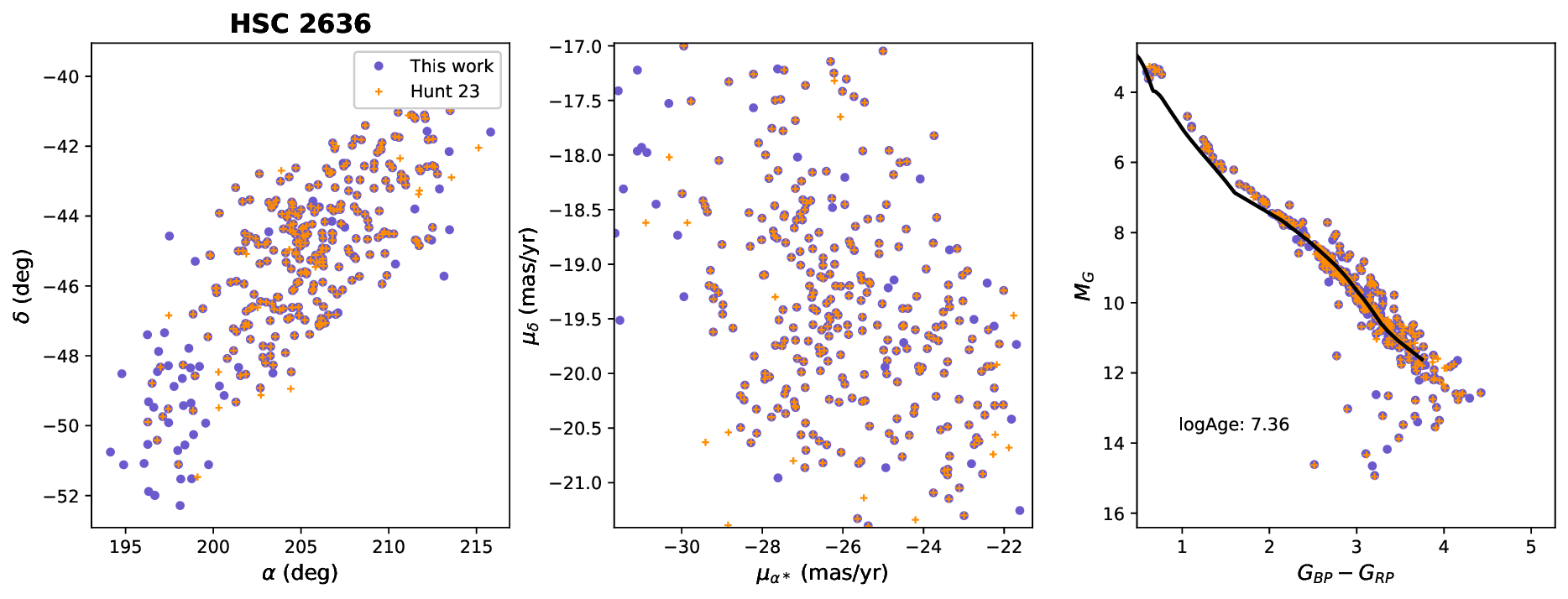}
}
\\[1.1cm]
{%
\includegraphics[width=0.9\textwidth, trim=0.1cm 0.9cm 0.0cm 1.6cm]{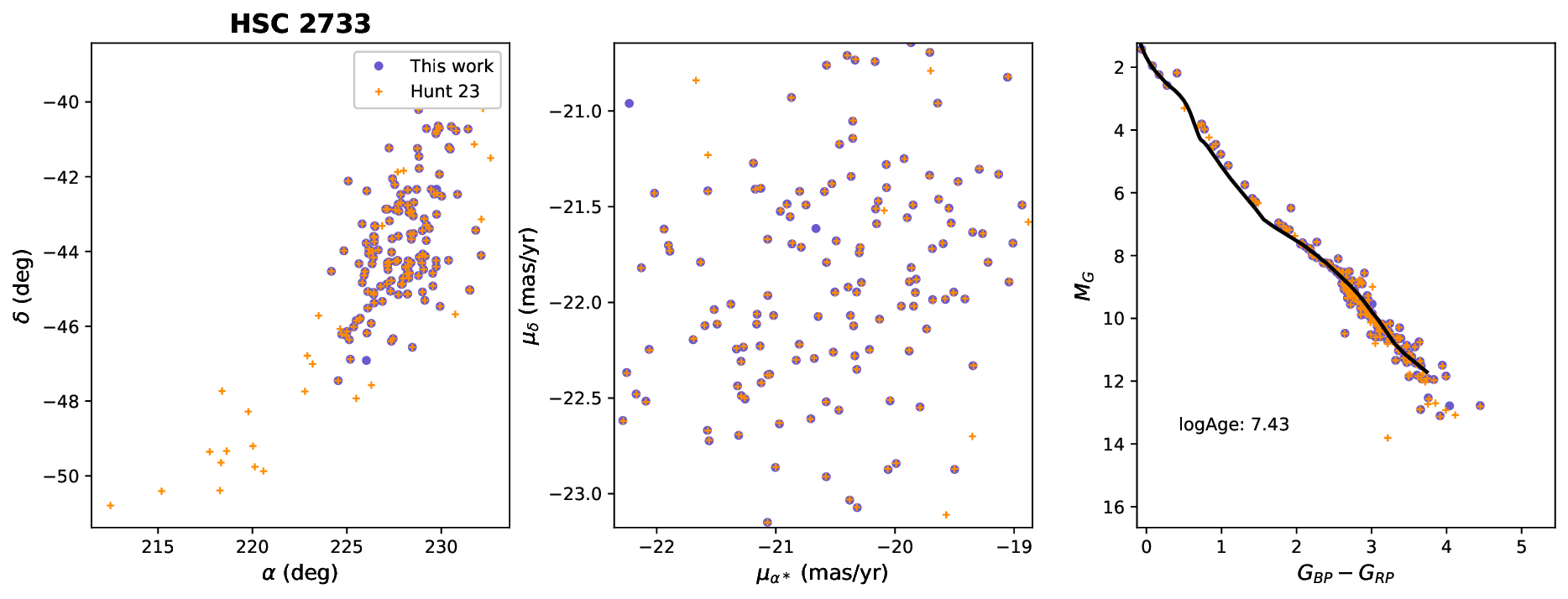}
}
\\[1.1cm]
{%
\includegraphics[width=0.9\textwidth, trim=0.1cm 0.9cm 0.0cm 1.6cm]{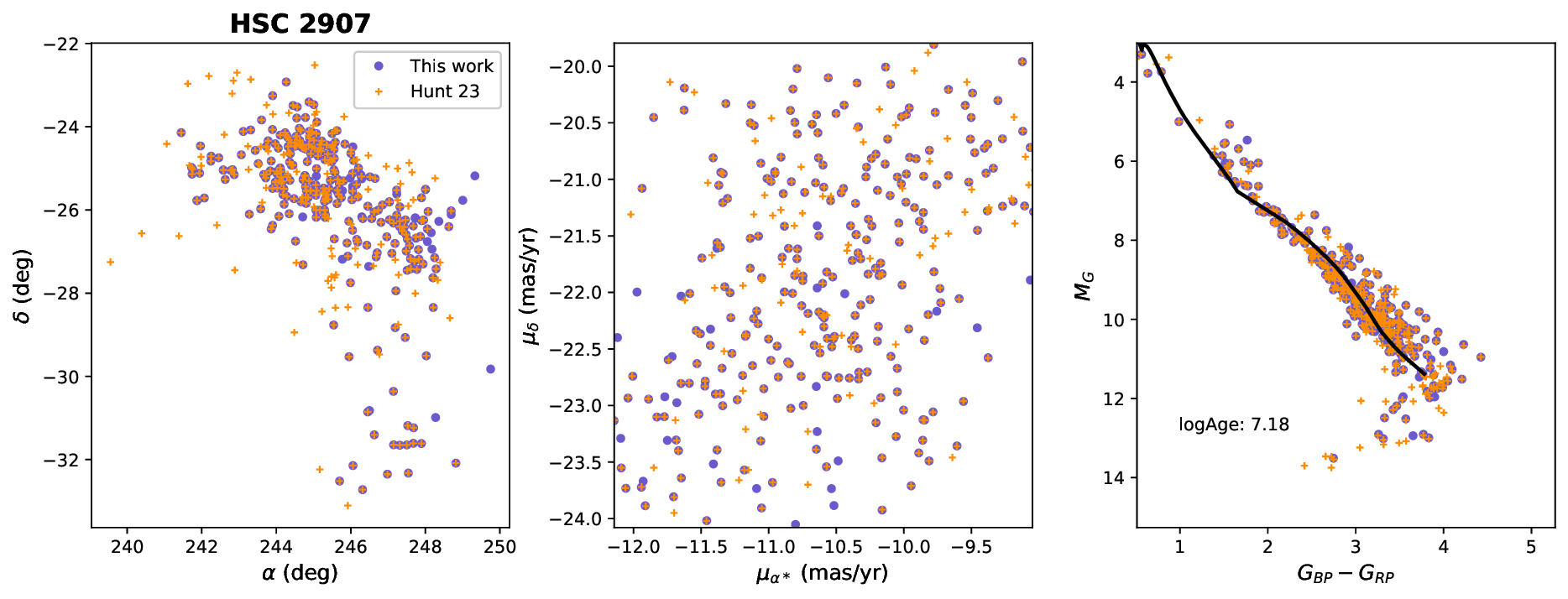}
}
\\[0.2cm]
\caption{Continued Fig~\ref{fig:apoc}}
\label{fig:apoc}
\end{figure*}

\setcounter{figure}{0}
\begin{figure*}
\centering
{%
\includegraphics[width=0.9\textwidth, trim=0.1cm 0.9cm 0.0cm 1.6cm]{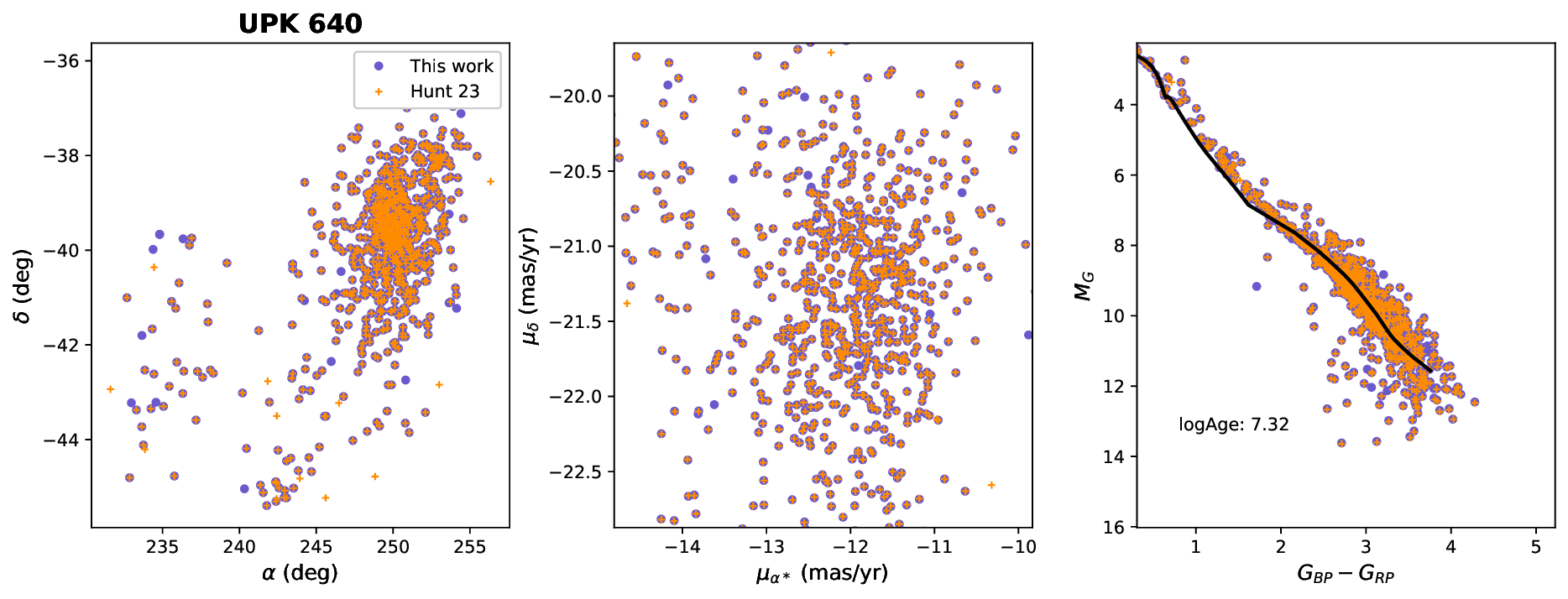}
}
\\[1.1cm]
{%
\includegraphics[width=0.9\textwidth, trim=0.1cm 0.9cm 0.0cm 1.6cm]{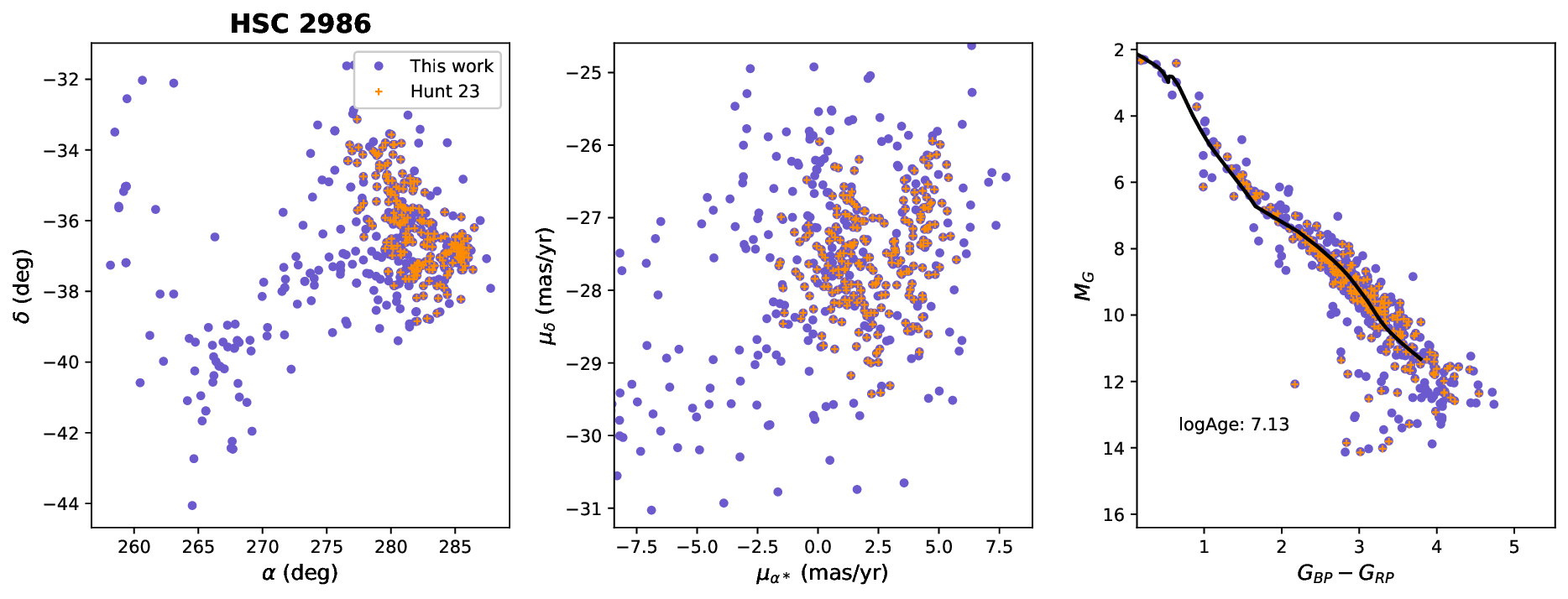}
}
\\[1.1cm]
{%
\includegraphics[width=0.9\textwidth, trim=0.1cm 0.9cm 0.0cm 1.6cm]{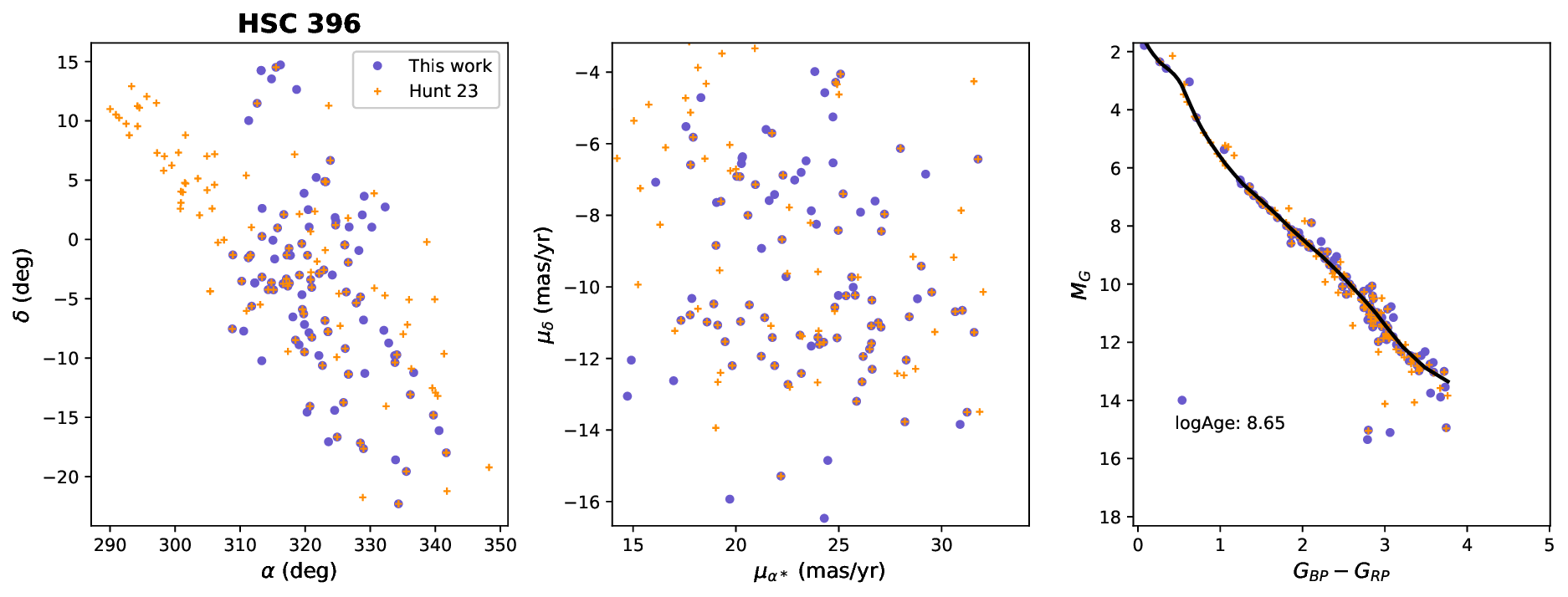}
}
\\[1.1cm]
{%
\includegraphics[width=0.9\textwidth, trim=0.1cm 0.9cm 0.0cm 1.6cm]{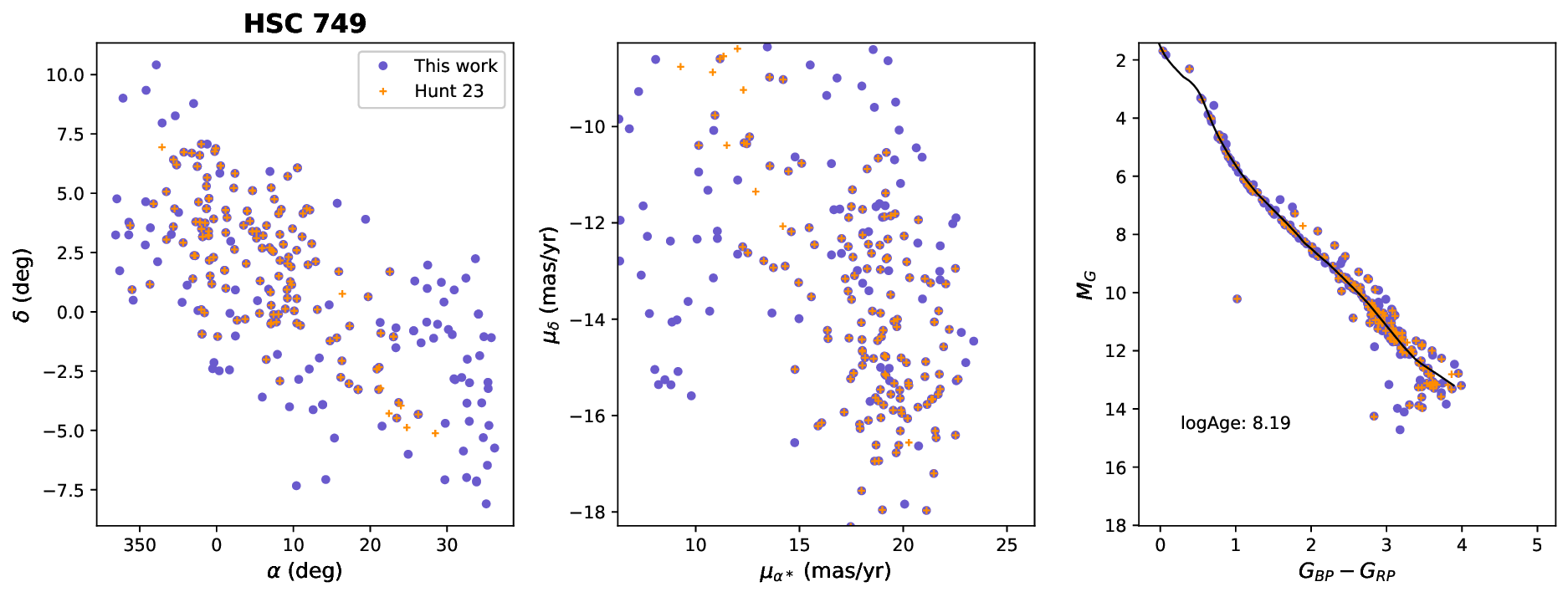}
}
\\[0.2cm]
\caption{Continued Fig~\ref{fig:apoc}}
\label{fig:apoc}
\end{figure*}

\setcounter{figure}{0}
\begin{figure*}
\centering
{%
\includegraphics[width=0.9\textwidth, trim=0.1cm 0.9cm 0.0cm 1.6cm]{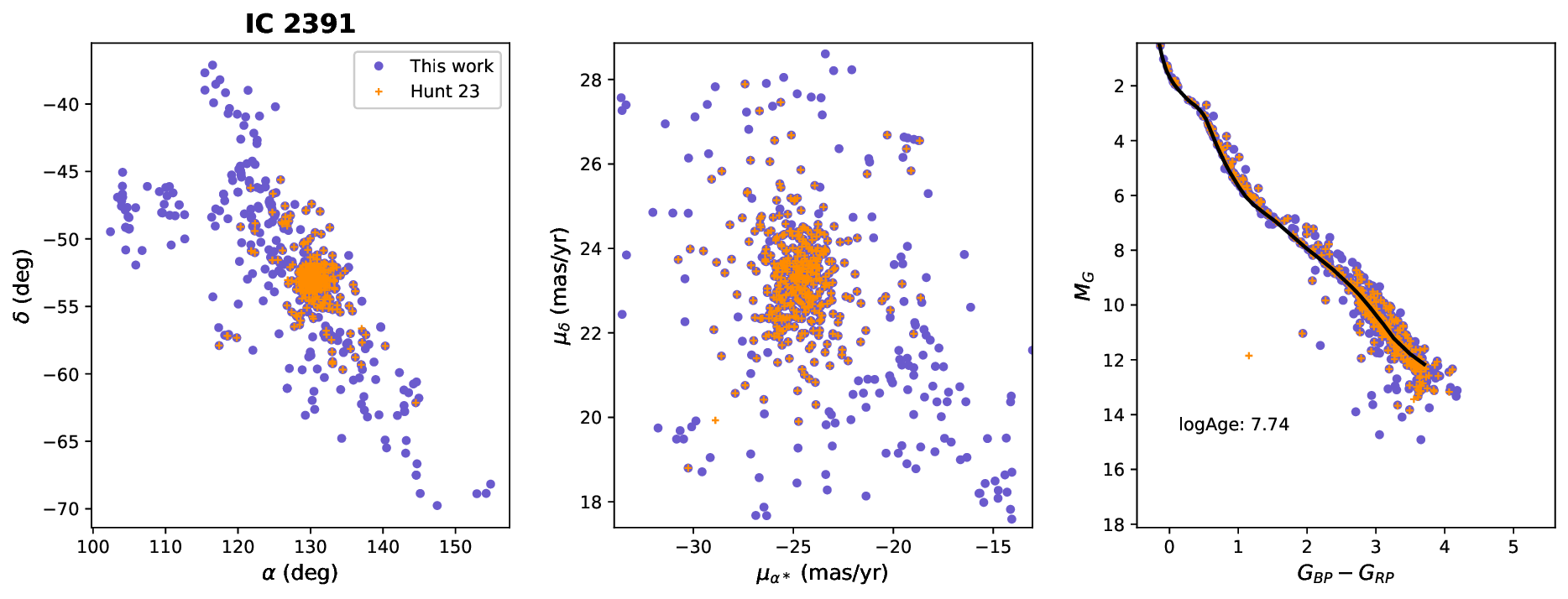}
}
\\[1.1cm]
{%
\includegraphics[width=0.9\textwidth, trim=0.1cm 0.9cm 0.0cm 1.6cm]{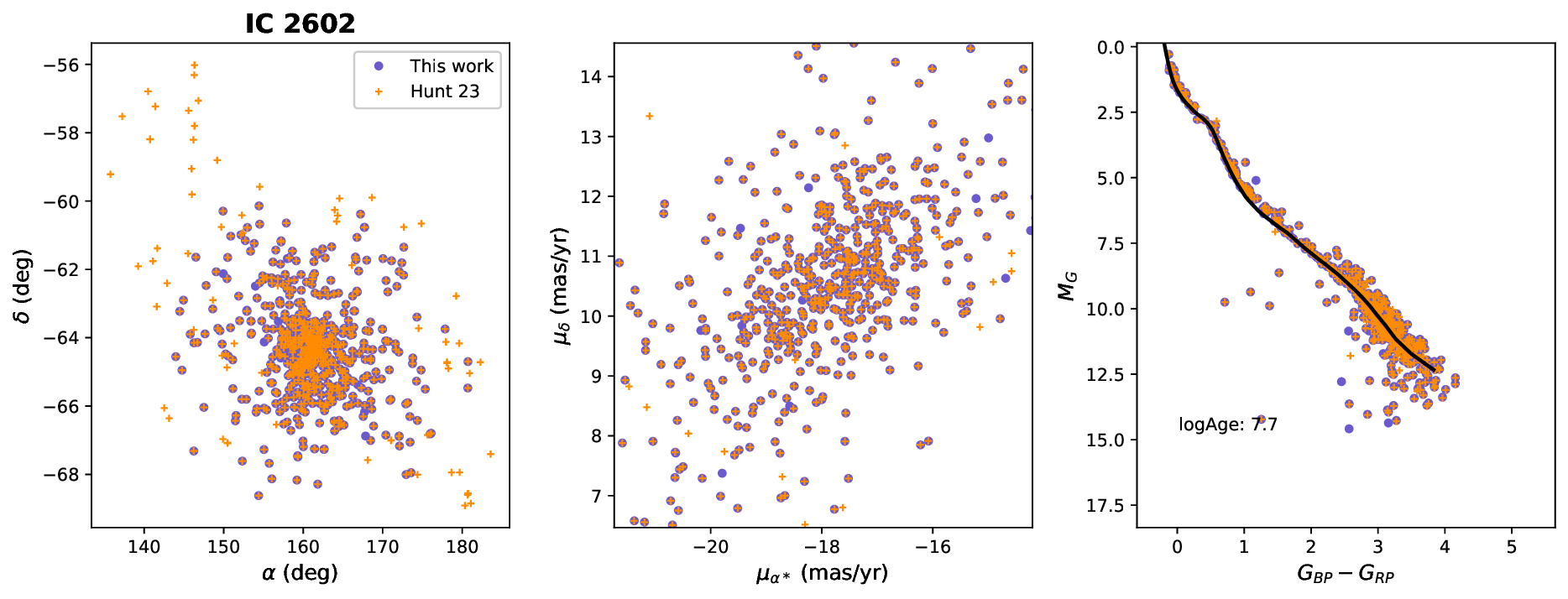}
}
\\[1.1cm]
{%
\includegraphics[width=0.9\textwidth, trim=0.1cm 0.9cm 0.0cm 1.6cm]{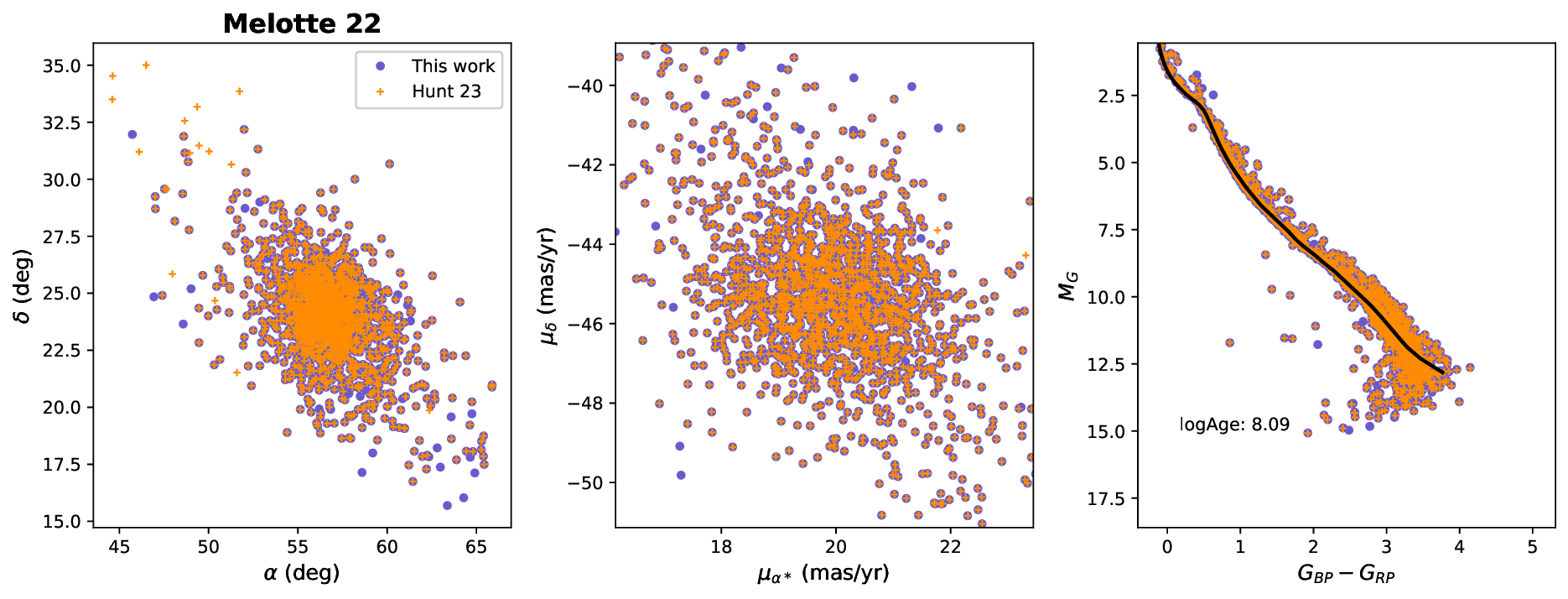}
}
\\[1.1cm]
{%
\includegraphics[width=0.9\textwidth, trim=0.1cm 0.9cm 0.0cm 1.6cm]{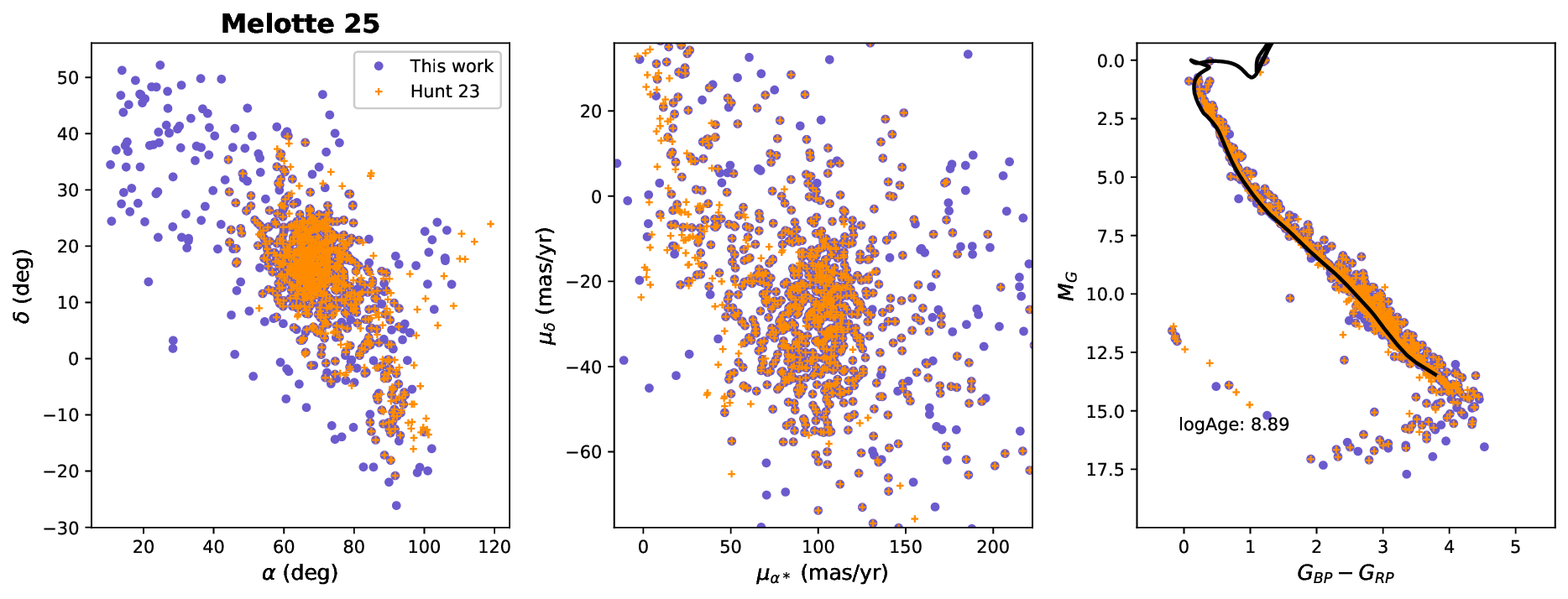}
}
\\[0.2cm]
\caption{Continued Fig~\ref{fig:apoc}}
\label{fig:apoc}
\end{figure*}

\setcounter{figure}{0}
\begin{figure*}
\centering
{%
\includegraphics[width=0.9\textwidth, trim=0.1cm 0.9cm 0.0cm 1.6cm]{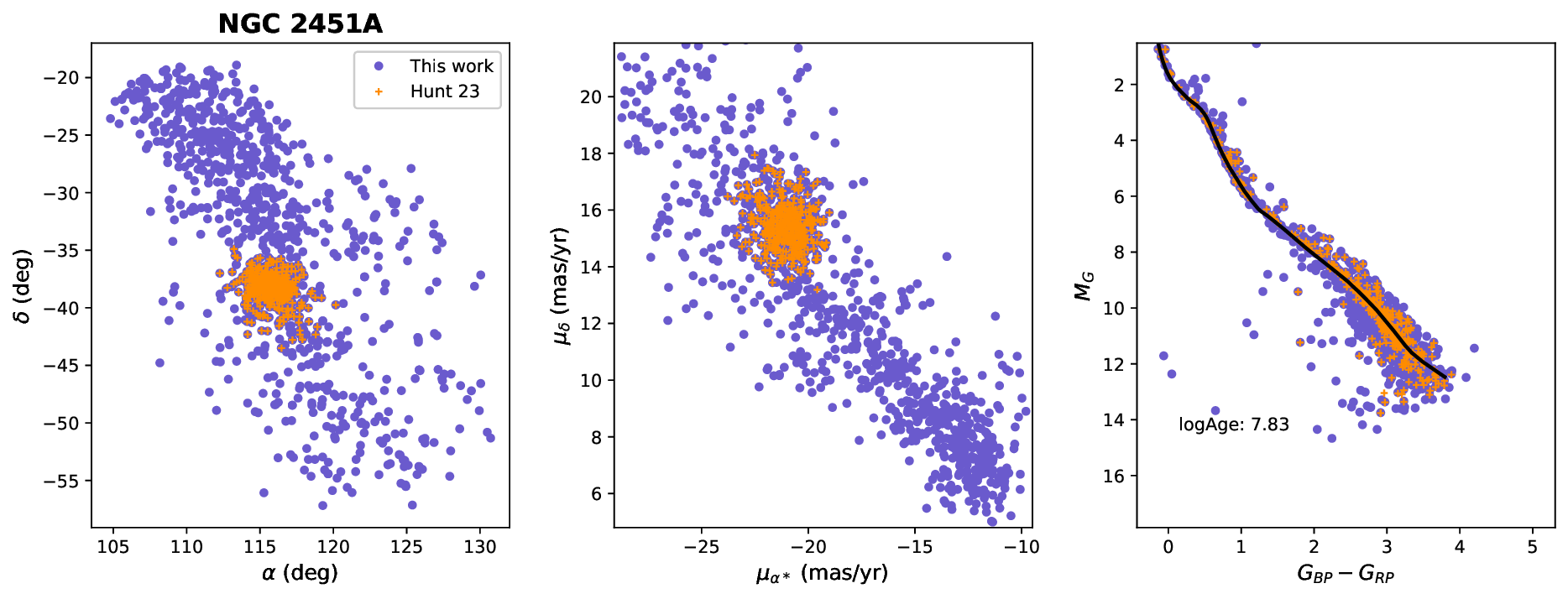}
}
\\[1.1cm]
{%
\includegraphics[width=0.9\textwidth, trim=0.1cm 0.9cm 0.0cm 1.6cm]{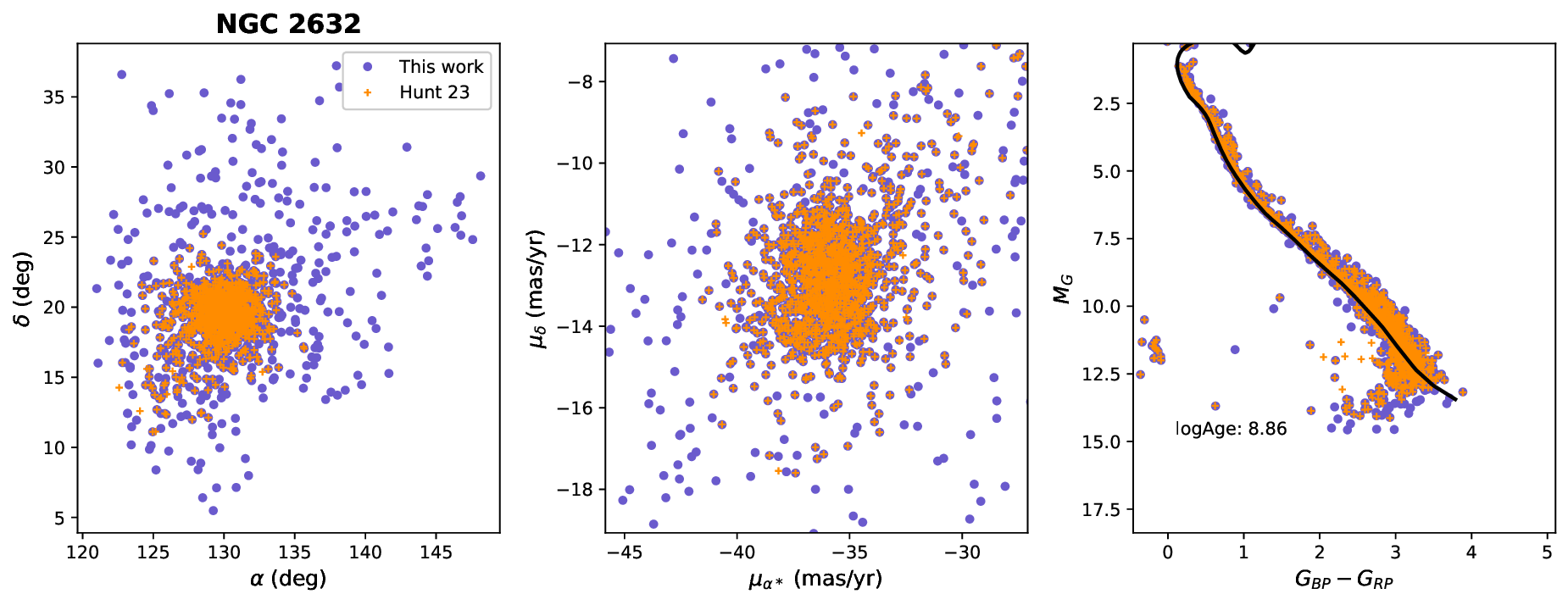}
}
\\[1.1cm]
{%
\includegraphics[width=0.9\textwidth, trim=0.1cm 0.9cm 0.0cm 1.6cm]{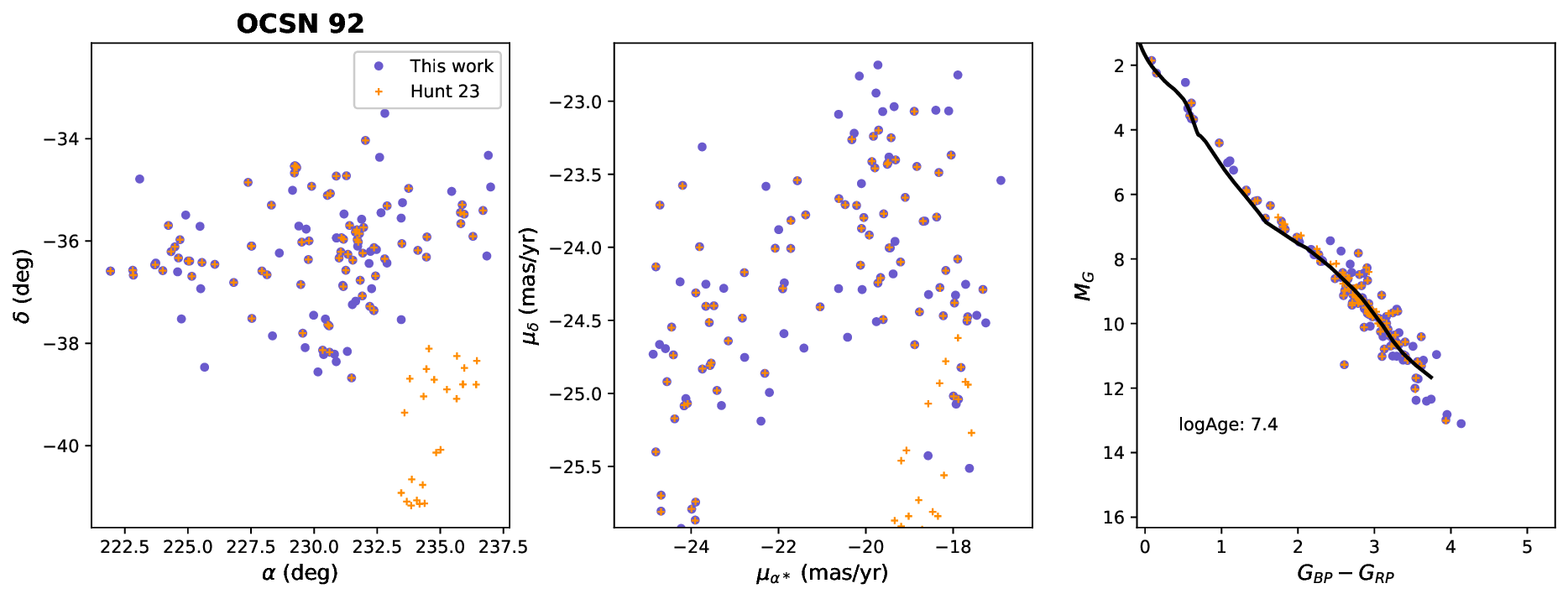}
}
\\[1.1cm]
{%
\includegraphics[width=0.9\textwidth, trim=0.1cm 0.9cm 0.0cm 1.6cm]{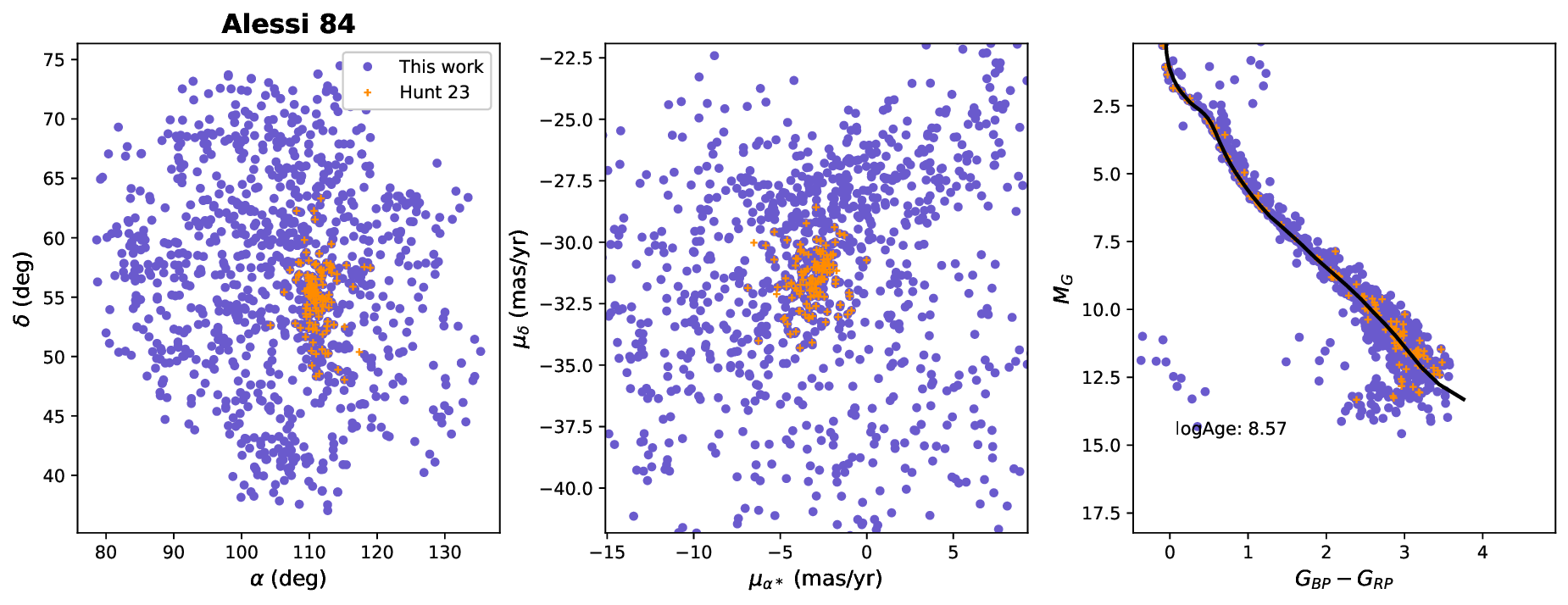}
}
\\[0.2cm]
\caption{Continued Fig~\ref{fig:apoc}}
\label{fig:apoc}
\end{figure*}

\setcounter{figure}{0}
\begin{figure*}
\centering
{%
\includegraphics[width=0.9\textwidth, trim=0.1cm 0.9cm 0.0cm 1.6cm]{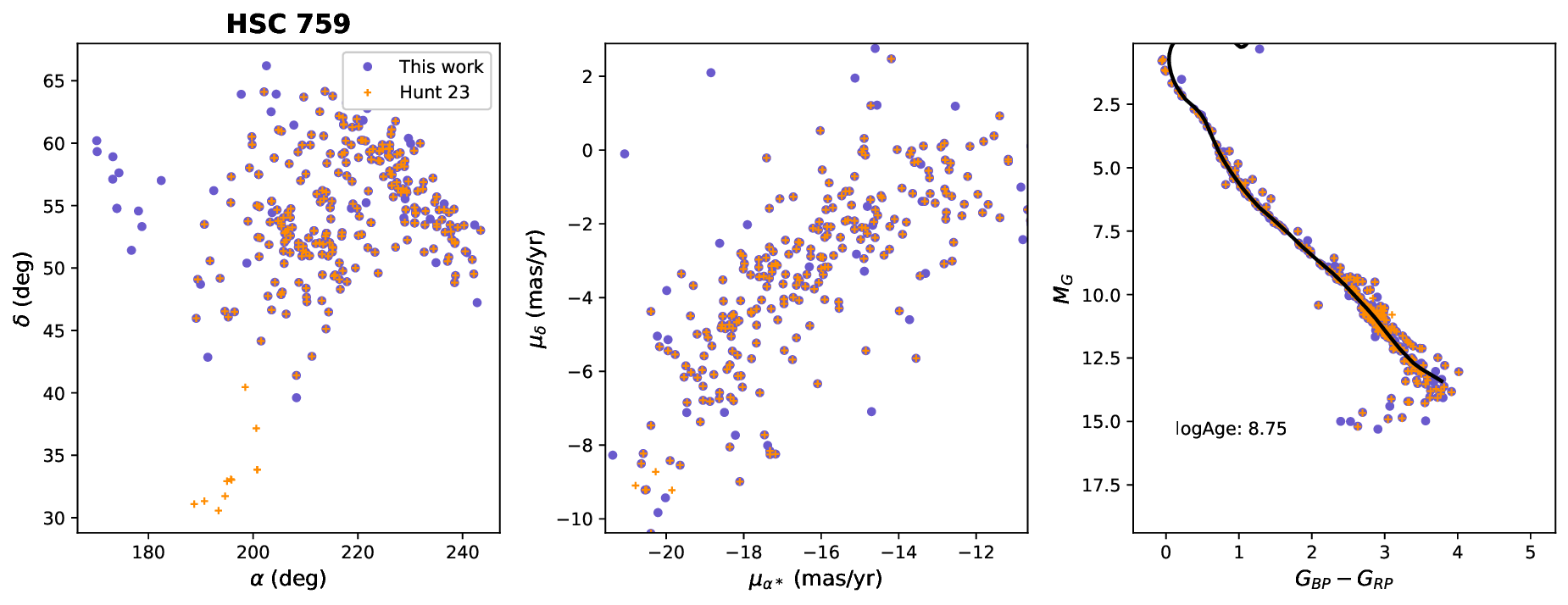}
}
\\[1.1cm]
{%
\includegraphics[width=0.9\textwidth, trim=0.1cm 0.9cm 0.0cm 1.6cm]{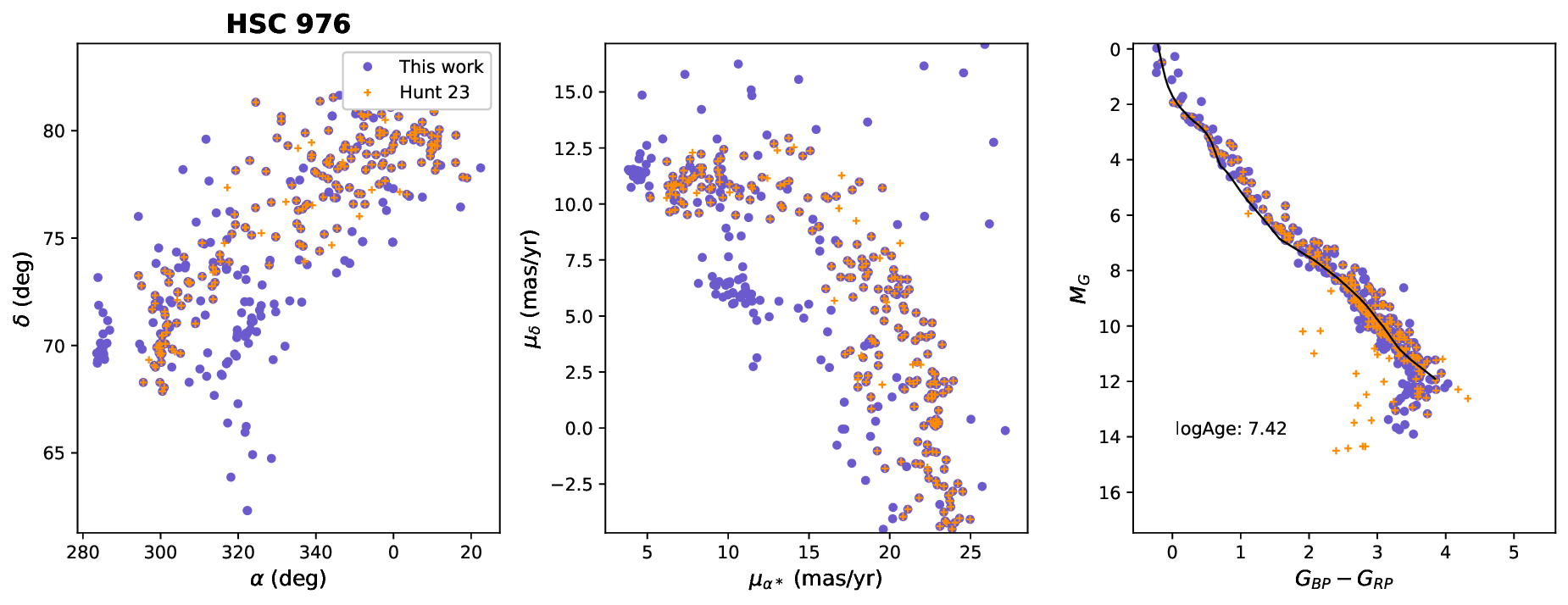}
}
\\[1.1cm]
{%
\includegraphics[width=0.9\textwidth, trim=0.1cm 0.9cm 0.0cm 1.6cm]{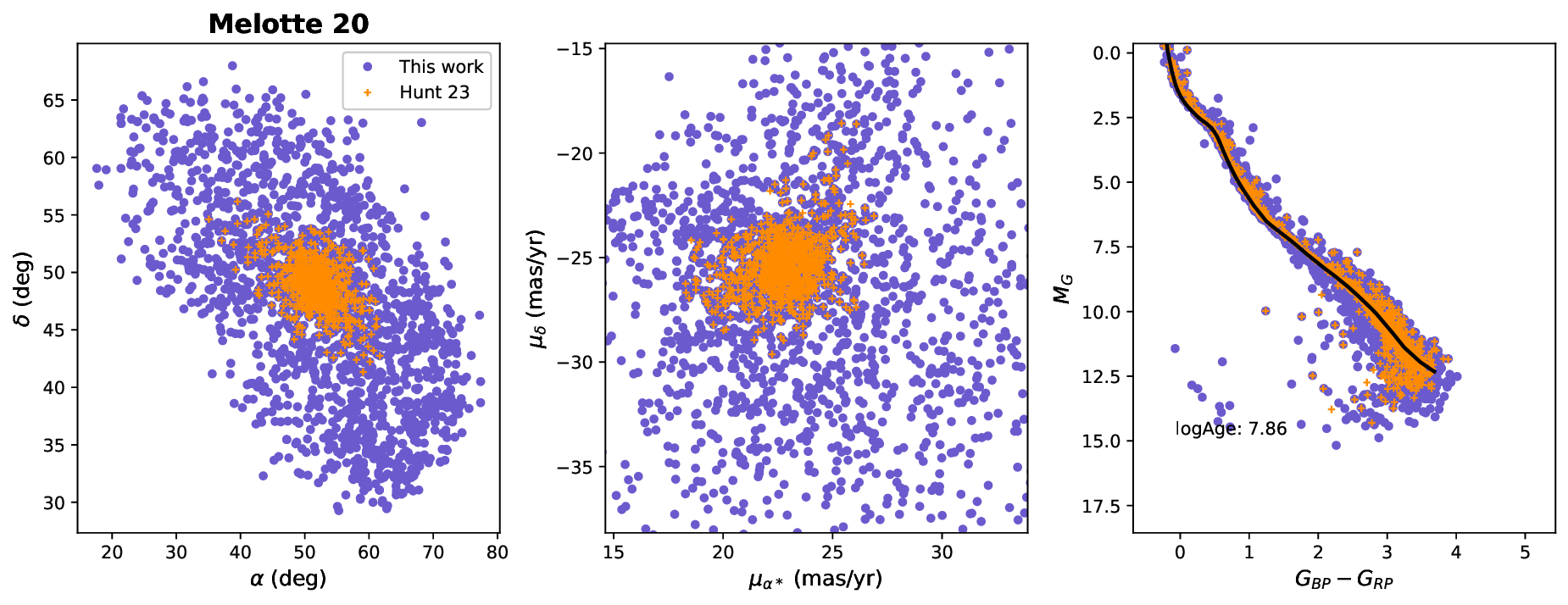}
}
\\[1.1cm]
{%
\includegraphics[width=0.9\textwidth, trim=0.1cm 0.9cm 0.0cm 1.6cm]{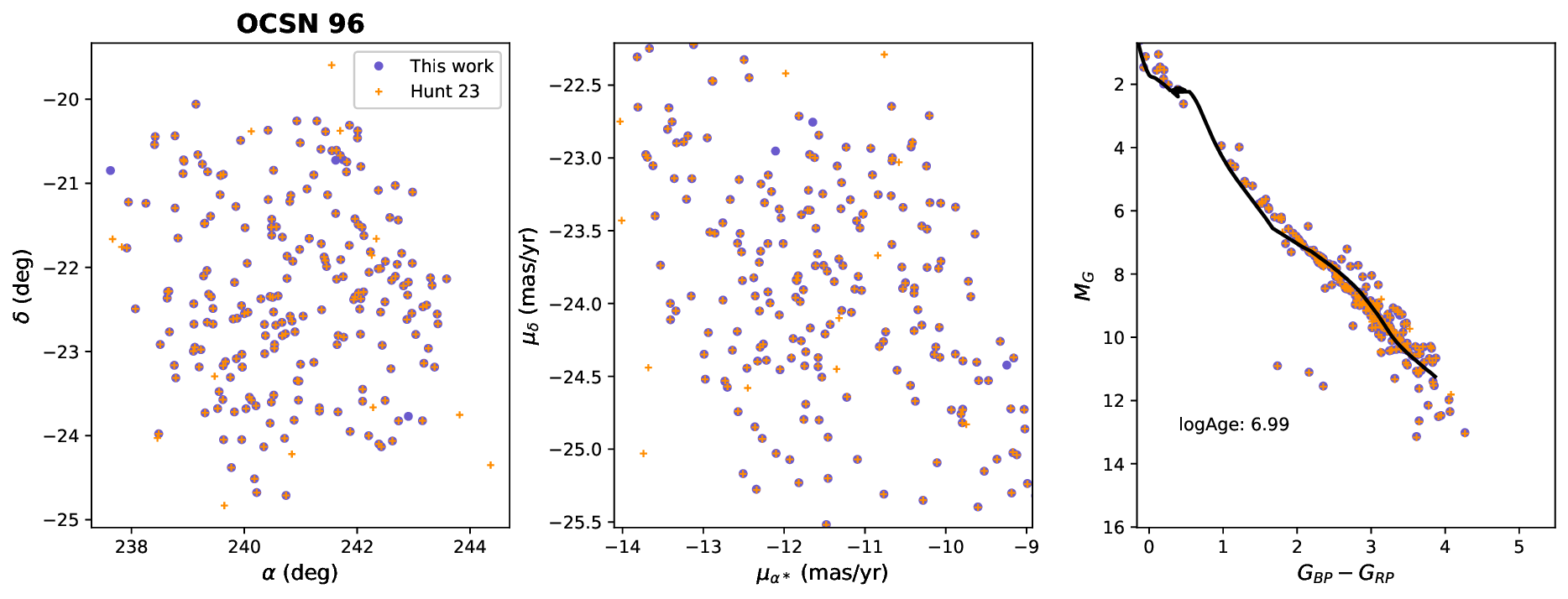}
}
\\[0.2cm]
\caption{Continued Fig~\ref{fig:apoc}}
\label{fig:apoc}
\end{figure*}

\setcounter{figure}{0}
\begin{figure*}
\centering
{%
\includegraphics[width=0.9\textwidth, trim=0.1cm 0.9cm 0.0cm 1.6cm]{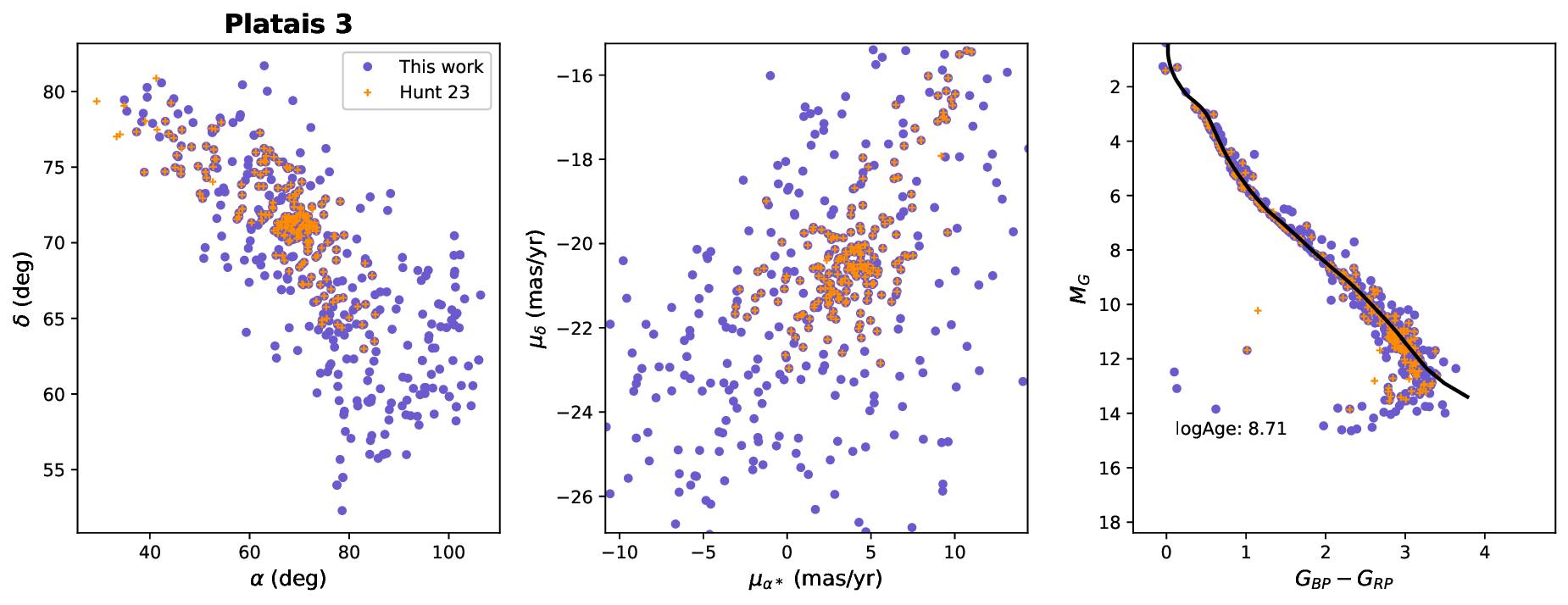}
}
\\[1.1cm]
{%
\includegraphics[width=0.9\textwidth, trim=0.1cm 0.9cm 0.0cm 1.6cm]{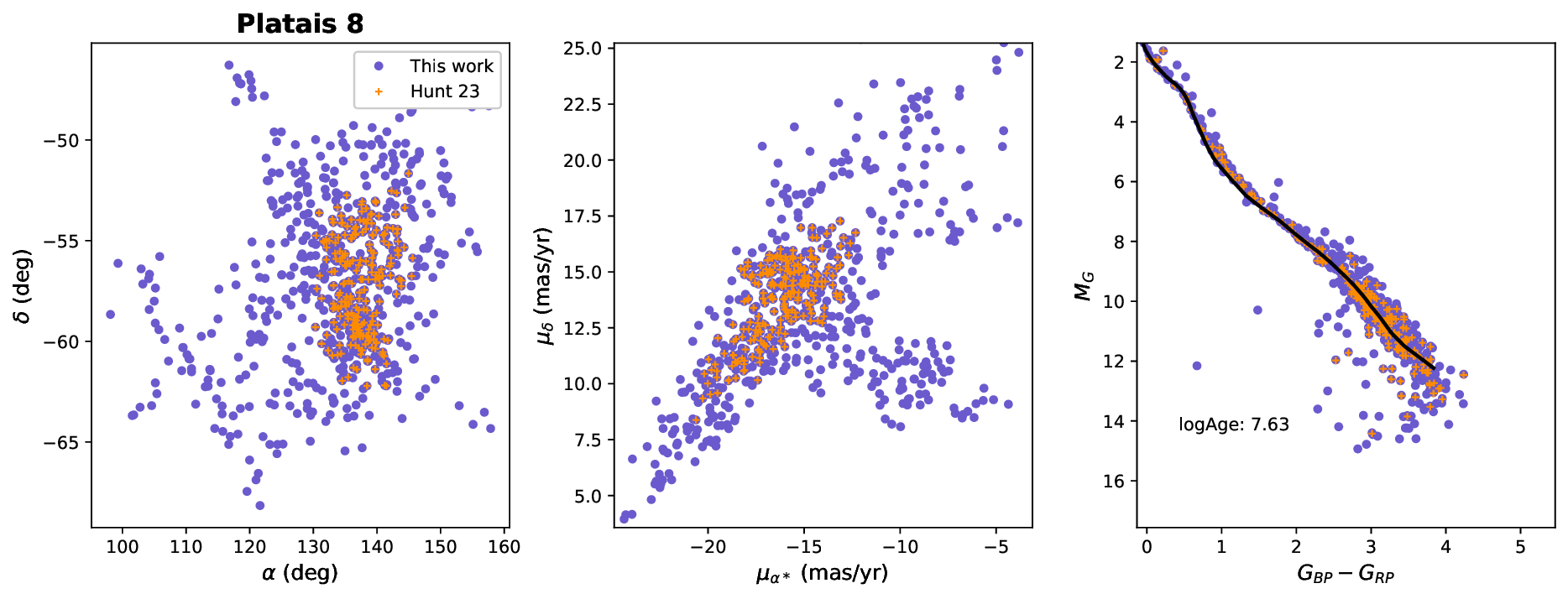}
}
\\[1.1cm]
{%
\includegraphics[width=0.9\textwidth, trim=0.1cm 0.9cm 0.0cm 1.6cm]{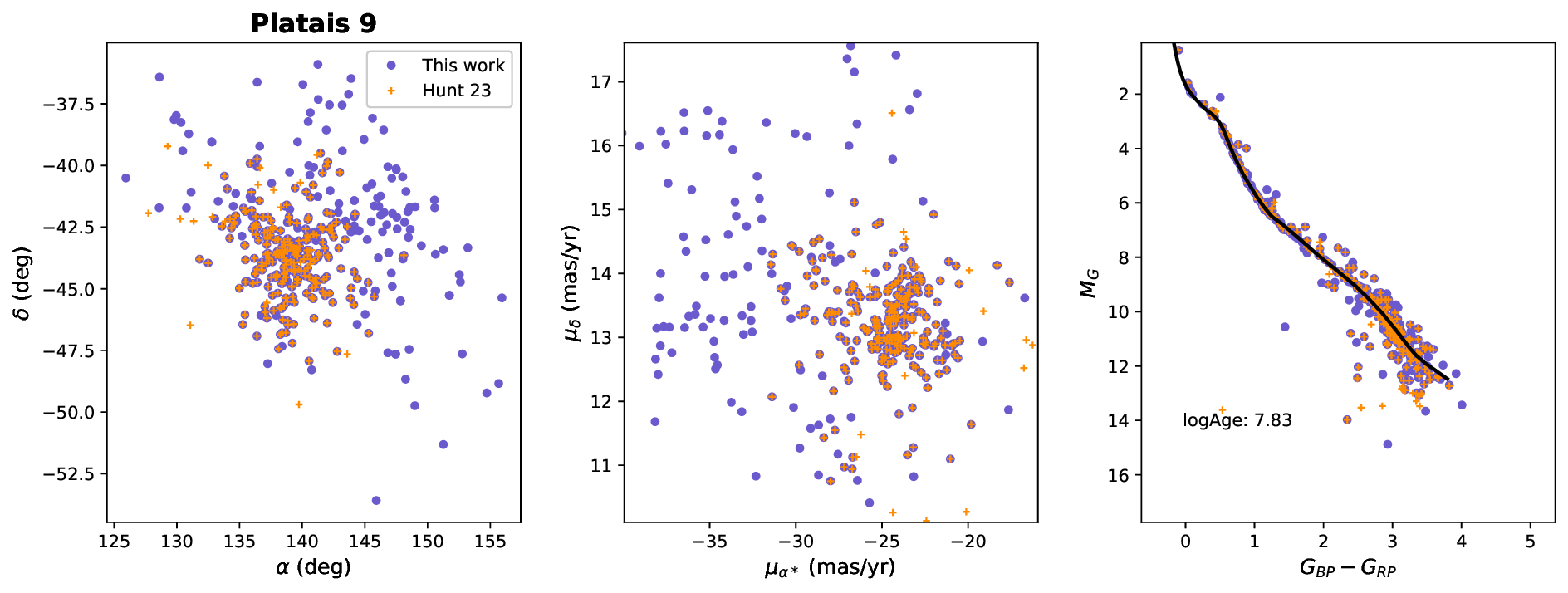}
}
\\[0.2cm]
\caption{Continued Fig~\ref{fig:apoc}}
\label{fig:apoc}
\end{figure*}

\label{app:appendix}
\end{appendix}

\end{document}